\documentclass[
notitlepage,
onecolumn,
nofootinbib,
amsmath,amssymb,aps, floatfix]{revtex4-1}

\usepackage{graphicx}
\usepackage{bm}
\usepackage{ulem}

\date{\today}
\usepackage{color}
\newcommand{\red}[1]{{\color{black} #1}}
\newcommand{\blue}[1]{{\color{black} #1}}
\newcommand{\green}[1]{{\color{black} #1}}

\begin{document}

\title{Characteristics of anomalous deterministic transport\\
in steady plane viscous flows\\ 
with stagnation points}

\author{Michael A. Zaks}
\affiliation{Humboldt University of Berlin, Berlin, Germany}

\author{Alexander Nepomnyashchy}
\affiliation{Technion - Israel Institute of Technology, Haifa, Israel}
\nopagebreak
\maketitle
\nopagebreak

\section{Introduction}
\label{sect:intro}
Fluid motions are often interesting and useful not only in themselves but also 
from the point of view of transport of particles carried by these fluids. 
To an observer who follows the ensemble of material particles advected by the fluid,
the temporal evolution of ensemble characteristics may look
simple or complicated, depending on the velocity field that advects the tracers. Remarkably, non-trivial transport is not necessarily
a consequence of turbulent or chaotic Eulerian dynamics: Lagrangian chaos \textit{aka} chaotic 
advection delivers numerous examples of aperiodic tracer motions in stationary (that is, time-independent) velocity fields~\cite{Aref-84,Ottino-89}. 

For passive particles carried by the fluid, inhomogeneities
of the velocity field result in local variations of the motion
characteristics. When an ensemble of drifting tracers traverses
a region of strong inhomogeneity, 
some of its elements are subjerevtex-1cted to accelerations, whereas the other ones experience slowdowns. 
In his seminal paper on diffusion
in random velocity fields~\cite{Kraichnan},
Robert Kraichnan, the classic of the modern turbulent research,
noted:  
\textit{``The simulation run in two dimensions ... shows anomalies
which appear to be clearly associated with trapping''} 
observing that those trapping effects were localized around the generic
maxima and minima of the stream function in the plane velocity fields. 
From the point of view of the velocity field, 
the maxima and minima of the stream function are stagnation
points in which both components of the velocity vanish.
In the purely deterministic context neither immediate vicinities of maxima nor those of minima
are visited by generic passive tracers: these stagnation points are \textit{elliptic}; 
in the flow pattern they are surrounded by
nested closed streamlines, impenetrable for the tracers.
This geometric localization puts maxima and minima in contrast
to the vicinities of the other important kind of singularities
of the stream function: \textit{hyperbolic} points, the saddles.
Tracers approach the saddle points along the incoming separatrices and
leave their vicinities along the outgoing ones. 
Globally, stagnation points of different kinds can coexist within the flow pattern. 
If the flow depends on external parameters (e.g., viscosity or the amplitude 
of the driving force), a variation of these parameters can create or destroy
stagnation points: from the point of view of dynamics, such events
are bifurcations of singular points in area-preserving plane flows. 
At bifurcation parameter values, the stagnation points, participating in the
bifurcation, are structurally unstable: degenerate.

\begin{figure}
\centering
\includegraphics[width=.5\linewidth]{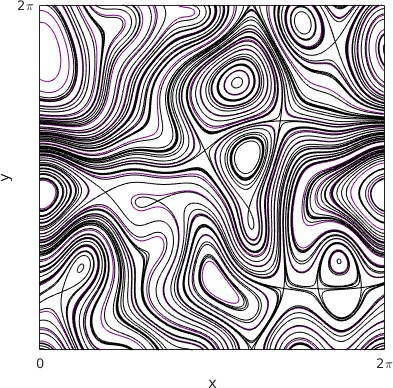}
\caption{Plane steady velocity field with periodic boundary conditions, 
stagnation points, confined eddies and unbounded jets. 
The field has been generated by the time-independent spatially periodic force
with Fourier harmonics up to the 4th order. The Figure is taken from~\cite{Kogler}}.
\label{fig:pattern}
\end{figure}

Fig.~\ref{fig:pattern} shows a characteristic stationary flow pattern
with eddies that are filled by continua of closed streamlines
and are encircled by separatrices of saddle stagnation points. 
Space between the eddies is filled by unbounded streamlines (``jets'').
If the flow pattern includes stagnation points, 
in their vicinities the fluid particles move significantly slower
than the rest of the ensemble, nearly coming to a halt; 
by the time when a temporarily trapped tracer finally leaves the neighborhood of the saddle, 
other particles are already drifting far downstream.
As a consequence, an initially compact cloud of tracers changes its shape and becomes
more and more stretched in the flow direction.
If the geometry of the flow ensures repeated passages of tracers 
across the zones of fast and slow motions, 
the net effect of passages affects the transport properties. 
Conventionally, transport of an ensemble is characterized in terms 
of the moments of the particle distribution, starting with the variance 
(sometimes the word ``dispersion'' is used):
\begin{equation}
\label{variance}
\mbox{var}(t)=\left\langle \big(x(t)-\langle x(t)\rangle\big)^2\right\rangle,
\end{equation}
the mean squared deviation of the appropriate coordinate $x(t)$ from the drifting center
of mass of the ensemble in the laboratory reference frame.
Below, we restrict ourselves to stationary two-dimensional (plane) fluid motions in rectangular domains
with periodic boundary conditions for velocity and pressure; there, the movement of a tracer is equivalent
to the motion upon the surface of the 2-torus and can be characterized e.g. by the rotation number.
We assume that the fluid is incompressible;
this ensures that the flow can in a standard way be described by its stream function and that the
fluid particles move along the streamlines: isolines of stream function.
We neglect molecular diffusion so that the dynamical system that governs the tracer motion
is completely deterministic. 

Although in the deterministic setup with its total absence of ``normal diffusion'' 
there is little sense in talking about  ``normal transport'', 
below we use the conventional term ``anomalous transport'' for the situations
in which the dependence of variance on time is different from linear.
Accordingly, we denote the sublinear and higher-than-linear dependencies as,
respectively, subdiffusive and superdiffusive transport.
We shall also consider the higher moments of the instantaneous distribution of tracers:
$\left\langle \big(x(t)-\langle x(t)\rangle\big)^n\right\rangle$
with $n>2$.

Direct calculations of the temporal evolution of variance for ensembles of tracers
carried by steady plane flows disclose the qualitative dependence of the growth
law on presence and the types of stagnation points in the flow pattern.
If such points are lacking, an originally compact ensemble of tracers remains compact forever, whereas variance stays bounded and oscillates at the low level.
If the flow pattern includes stagnation points,
variance in the transported ensemble displays unbounded growth.
If such points are hyperbolic (saddles), the time growth
of variance is approximately logarithmic, whereas the
presence of degenerate stagnation points enhances the growth, 
turning its time dependence into a power law:
the higher the degree of degeneracy, the larger the power.
The time of passage of a tracer across the flow region with a stagnation point diverges
when at the entrance to that region the tracer approaches the incoming separatrix of that point. 
In an earlier paper~\cite{pnas} we related the divergence rate to the growth rate
of variance in an ensemble of advected tracers, providing explicit quantitative estimates
for several typical situations. 
We distinguished between the logarithmic and the power-law singularities of passage time,
and in the latter group between weaker (leading to subdiffusive transport) 
and stronger (enabling superdiffusion) singularities.  

In the text below, we explain the derivation of the estimates from~\cite{pnas}:
we analyze the impact of stagnation points on the variance and higher moments of the tracer distribution
 for stationary, two-dimensional, spatially periodic flows.
In Sect.~\ref{ref:sect_hydr} we introduce the exemplary family
of stationary solutions of the forced Navier-Stokes equations. 
We discuss the basic patterns of streamlines for this family of flows, 
perform direct studies of transport processes, and present numerical evidence
that in the ensemble of tracers carried by the flow, 
the temporal evolution of the variance depends on the
presence/absence of stagnation points in the flow pattern. 

We also calculate the quantitative characteristics of transport. 
These empirical findings serve as a base for theoretical treatment in further sections.
In Section~\ref{sect:passage_time} we establish the explicit quantitative relation
between the singularities of passage time and the local characteristics
of the vector field at stagnation points. 
Further in that section we reformulate the problem in terms of the
\textit{special flow}, a convenient model that combines the properties of 
time-continuous flows and discrete mappings.
In Section~\ref{sect:numeric}, we \blue{focus on numerical studies of special flows
which model several hydrodynamically relevant situations.
For these special flows, we determine the patterns of temporal evolution 
of the variance, 
and discuss the role of rare events in approaching the asymptotic laws.} 
In Section~\ref{sect:theory}, we consider two simplified models of the phenomenon
that allow us to evaluate the ``trends'' of these asymptotic laws. 
For weaker singularities of the passage time, we replace the distribution of points,
which is created by the special flow, with an equidistant set of points. For stronger singularities, 
we suggest a unidirectional continuous-time random walk model. 
In both cases, we evaluate the ``trends" of the temporal evolution of moments: 
quantitative characteristics of growth that do not significantly depend
on the particular value of the irrational rotation number. 
Section~\ref{sect:conclusions} offers conclusions and a discussion.

\section{Deterministic transport in steady flow patterns:\\ 
examples and phenomenology}
\label{ref:sect_hydr}

\subsection{Case study: evolution of the flow pattern on a torus}
To illustrate the interrelation between the deterministic transport
of passive tracers and the underlying flow pattern, we employ
the exemplary family of plane steady incompressible flows~\cite{ZPK_96}. 
The flow is induced by the time-independent force 
on a square domain with periodic boundary conditions.
The Navier-Stokes equations that govern the fluid motion are 
\begin{eqnarray}
\label{navier}
\frac{\partial\bm v}{\partial t}+ ( \bm v\cdot \nabla)\,\bm v  & = &
-\frac{ \nabla P}{\rho_0} + \nu\,\nabla^2 {\bm v}\;+\,\bm{F}\nonumber\\[-1.5ex]
& & \\[-1.5ex]
\nabla\cdot\bm v &=&0\nonumber
\end{eqnarray}
with $\bm{v}$ and $P$ being, respectively, velocity and pressure, 
whereas material constants $\rho_0$ and $\nu$ denote the density 
and the kinematic viscosity of the fluid.
We align the coordinate axes with the square sides,
set the side length at $2\pi$ 
and choose the doubly periodic forcing term: 
${\bm F} = (f\sin y,f\sin x,0)$, with $f$ being the constant
forcing amplitude; this is a kind of generalization for the
Kolmogorov flow~\cite{Arnold-Meshalkin-60,Meshalkin-Sinai-61}
in which the force acts along one of the coordinate axes
and is spatially periodic with respect to the complementary coordinate.
On imposing spatial periodicity for the velocity field,
\begin{equation}
\label{bc1}
\bm{v}(x,y)=\bm{v}(x+2\pi,y)=\bm{v}(x,y+2\pi),
\end{equation}
 the original plane domain becomes equivalent to the surface of the 2-torus.

Being interested in the long-range advection, 
we prescribe constant nonzero mean drift across the domain
in both spatial directions, parameterizing it by two flow rates $\alpha$ and
$\beta$, respectively:
\begin{equation}
\label{bc2}
\frac{1}{2\pi}\int_0^{2\pi}v_x dy 
=\alpha,\qquad
\frac{1}{2\pi}\int_0^{2\pi}v_y dx=\beta .
\end{equation}
Fixed mean drift can be achieved e.g. by constant uniform pumping along the
coordinates $x$ and $y$.
Incompressibility of the fluid allows us to 
introduce the stream function $\Psi(x,y):\;
v_x=\partial\Psi/\partial y,\;\;
v_y=-\partial\Psi/\partial x$. 
This step casts the problem into the Hamiltonian frame,
in which two components of velocity are canonical
variables, and $\Psi$ plays the role of energy.
In terms of stream function, the forced Navier-Stokes equations (\ref{navier})
read
\begin{equation}
-\frac{\partial \Delta\Psi}{\partial t}+
\frac{\partial \Psi}{\partial x}
\frac{\partial \Delta\Psi}{\partial y}-
\frac{\partial \Psi}{\partial y}
\frac{\partial \Delta\Psi}{\partial x}\,=\,
-\nu \Delta\Delta\Psi + {\mbox{\textbf{curl}}}_z {\bm F}.
\label{Navier_psi}
\end{equation}

For the considered ${\bm F}$ the last term in (\ref{Navier_psi})
yields $f(\cos y - \cos x)$, and the equation
has the time-independent (steady) solution~\cite{ZPK_96}
\begin{equation}
\label{stat}
\Psi=\alpha y-\beta x + \frac{f\;\sin(x+\phi_1)}{\sqrt{\alpha^2+\nu^2}}
-\frac{f\;\sin(y+\phi_2)}{\sqrt{\beta^2+\nu^2}} 
\end{equation}
$$\mbox{with}\quad \phi_1=\arctan\frac{\nu}{\alpha},\quad\phi_2=\arctan \frac{\nu}{\beta}.$$
In a plane stationary velocity field, passive tracers move along the streamlines: lines of constant $\Psi$.
Equations of motion for advected particles are
\begin{equation}
\label{velocity_field}
\dot{x}=v_x=\alpha-\frac{f\;\cos(y+\phi_2)}{\sqrt{\beta^2+\nu^2}},\qquad
\dot{y}=v_y=\beta-\frac{f\;\cos(x+\phi_1)}{\sqrt{\alpha^2+\nu^2}}
\end{equation}

Stability of the stationary flow pattern~(\ref{velocity_field}) with respect to small perturbations
was numerically studied in~\cite{WZSS_2020} for the wide range of the values of
the forcing amplitude and pumping intensities.  
In the framework of the corresponding linearized equation, it was found that
all perturbations whose wavelengths were divisors of the size
of the square domain decayed.  
This allows us to restrict the studies of transport
in the flow described by the PDE (\ref{navier}) to the advection governed by
the second order ODE (\ref{velocity_field}).

Below we illustrate peculiarities of transport in this field
under various characteristic combinations of the parameters.
Aiming at motions of tracers across large distances, 
we lift the torus flow (\ref{velocity_field}) onto the plane. 
Upon the plane, the streamlines can be divided into bounded
(contained within a finite number of periodicity cells) and unbounded ones.
Under the fixed set of parameter values, all unbounded streamlines of (\ref{stat}) share 
the same mean inclination with respect to the coordinate axes; 
this is
\begin{equation}
\label{rho}
\rho=\lim_{t\to\infty}\,\frac{x(t)-x(0)}{y(t)-y(0)}
\end{equation}
where the starting point $\left(x(0),y(0)\right)$ 
lies anywhere on any unbounded streamline: 
the \textit{rotation number} on the torus,
the fundamental global characteristics of the torus dynamics. 
When, like in the case of (\ref{stat}), the stream function is a superposition
of the linear and periodic functions of the coordinates, 
$\Psi(x,y) =\alpha y-\beta x +\Phi(x,y)$, with $\Phi(x,y)$ being periodic
in both arguments and having the same period length for both, the
rotation number is determined by the linear part alone: 
\begin{equation}
\label{rho_lin}
\rho=\frac{\alpha}{\beta}.
\end{equation}
From the point of view of the original setup, the rotation number is the ratio of two
mean drift rates, in other words, of two pumping intensities. By freezing these intensities 
and varying the forcing amplitude $f$ (and, if necessary, the viscosity $\nu$) 
we can follow the changes in the flow pattern and in the transport characteristics 
under the fixed rotation number. Before proceeding in that direction, an important remark:
All unbounded streamlines, if restricted to the torus surface,
are closed under rational values of $\rho$ and non-closed under irrational ones.
In the former case the motion of every tracer particle is periodic; notably, the length
of the period depends on the chosen streamline.

\begin{figure}[h]
\centering
\includegraphics[width=.23\linewidth]{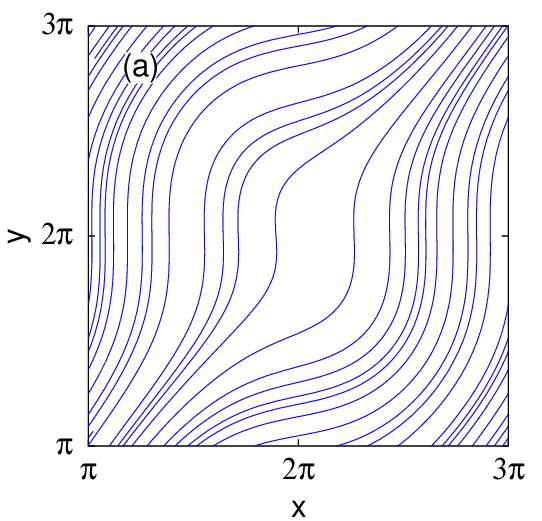}\quad
\includegraphics[width=.23\linewidth]{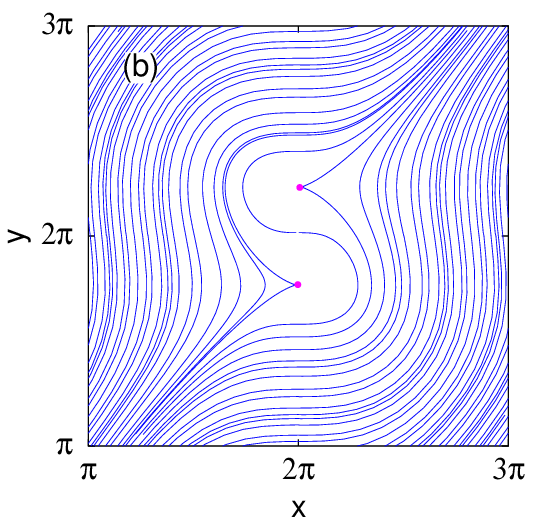}\quad
\includegraphics[width=.23\linewidth]{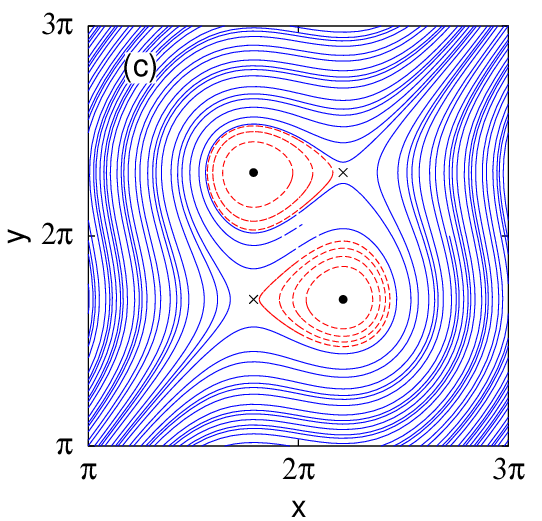}\quad
\includegraphics[width=.23\linewidth]{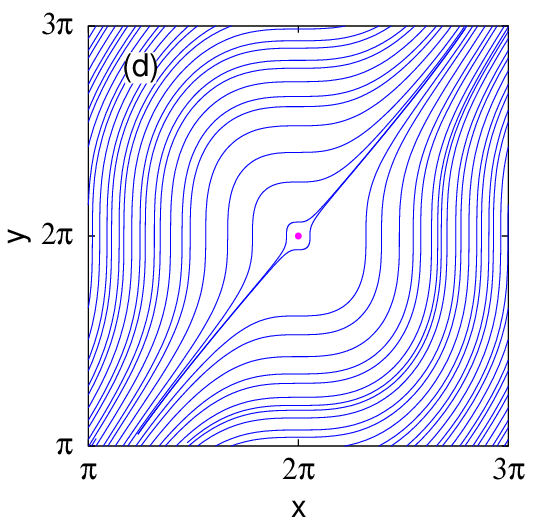}
\caption{Flow patterns, defined by Eq.(\ref{stat}) at $\alpha=(\sqrt{5}-1)/2,\,\beta=1$ \ 
(``golden mean'' rotation number) and different combinations of $f$ and $\nu$.\quad
(a) $\nu=1$, \ $f=0.9<f_{\rm cr}$; (b) $\nu=1$, $f=f_{\rm cr}=1.17557$;
(c) $\nu=1$, \ $f=1.5>f_{\rm cr}$; (d) $\nu=0$, $f=f_{\rm cr}=0.61803$.\\
Crosses: saddle stagnation points, filled circles: non-hyperbolic stagnation points;
dashed red lines in (c): closed streamlines.\\
Patterns (a) and (c) are robust; patterns (b) and (d) contain degenerate
stagnation points.
}
\label{fig:4patterns}
\end{figure}

In the absence of forcing ($f=0$) the 
streamlines are straight; velocity is uniform everywhere on the torus.
An increase of $|f|$ makes velocity depend on the
coordinates, so that the streamlines become curvy (Fig.~\ref{fig:4patterns}a). 
Within the interval of the force amplitude
$0<|f|<f_{\rm cr}=\sqrt{\alpha^2\beta^2+\nu^2\max(\alpha^2,\beta^2)}$,
the motion of a tracer carried by this flow is periodic for rational $\rho$ and quasiperiodic
for irrational $\rho$; in both cases the Fourier spectra of generic Lagrangian observables 
(e.g. of velocity components) are discrete. 

The value $|f|=f_{\rm cr}$ marks a transformation in the flow pattern: two turning points 
simultaneously appear upon two streamlines (Fig.~\ref{fig:4patterns}b);
in each of them both components of velocity vanish. Immediately beyond $f_{\rm cr}$, 
each of the turning points splits into a pair of stagnation points:
an elliptic point and a saddle stagnation point. 
These local events have global implications (Fig.~\ref{fig:4patterns}c): 
at $|f|>f_{\rm cr}$ two isolated stationary eddies are present 
on the background of mean drift; each vortex is centered 
at the corresponding elliptic point, 
is filled with a continuum of closed streamlines and is bounded
from the outside by the separatrix of the saddle stagnation point.
From the point of view of the Hamiltonian flow (\ref{velocity_field}), the transition
at $f_{\rm cr}$ is a (codimension-1) saddle-center bifurcation; beyond it, the phase
space becomes separated into the ``local'' component, filled by closed bounded streamlines
and the ``global'' component, which consists of unbounded streamlines.
If the rotation number is irrational, then in the global component
every streamline is dense on the torus and includes repeated passages
arbitrarily close to the stagnation point. For a tracer particle, epochs of
relatively fast motion at a distance from stagnation points alternate with epochs of long, nearly stagnant
hovering near the equilibria. Therefore, the motion past
stagnation points with adjacent eddies is more involved than the conventional quasiperiodicity:
Fourier spectra of Lagrangian observables are singular continuous
and are supported by fractal sets of frequency values~\cite{ZPK_96}.

A generic route from a flow pattern without eddies to a pattern with a single vortex
leads through the saddle-center bifurcation. 
In the family (\ref{velocity_field}) two such bifurcations
occur simultaneously at two different locations (Fig.~\ref{fig:4patterns}b),
leaving two counterrotating eddies in the flow pattern. 
In an inviscid flow ($\nu=0$) the degenerate variant of transition occurs: 
at $f_{\rm cr}=\alpha\beta$ a single non-hyperbolic equilibrium emerges 
on one of the streamlines (Fig.~\ref{fig:4patterns}d), and, following an arbitrary
small increase of $f$, evolves into the configuration, topologically equivalent to that  
of Fig.~\ref{fig:4patterns}(c):
four equilibrium points and two counterrotating
eddies filled with closed streamlines.

\subsection{Transport: empirical observations}
Consider now an ensemble of tracers, advected by the flow pattern (\ref{stat}).
We characterize transport in terms of the time-dependent variance of the ensemble, 
defined by Eq.(\ref{variance}). 
Without restricting generality, we use the coordinate $x$ for this characteristic.

At $f=0$ the velocity field is uniform
and an ensemble of tracers is merely translated
downstream without changes in size and shape;
hence the variance stays constant.

At $f\neq 0$ variations of velocity along the streamline lead to temporal fluctuations of variance. 
Here, the cases of rational and irrational values of the 
rotation number need to be distinguished.
If $\rho$ is rational, all streamlines on the torus are closed. For tracers sitting on
two different streamlines, the turnover times are, in general, different;
therefore, the average distances between the tracers grow linearly in time.
This trivial linear stretching of the ensemble should not be viewed 
as a proper ballistic transport because the pathways
of different tracer particles stay separated for all time values, 
and ensemble averaging does not make much sense.
Therefore, from now on we concentrate on irrational values of $\rho$.
In the absence of stagnation points in the flow pattern, that is, for $|f|<f_{\rm cr}$,
the time of passage across the square domain along every streamline is finite. 
As a consequence, variance in the tracer ensemble remains bounded at all times. 

\begin{figure}[h]
\centering
\includegraphics[width=.35\linewidth]{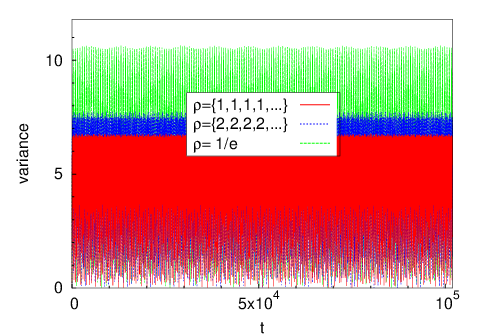}\quad
\includegraphics[width=.35\linewidth]{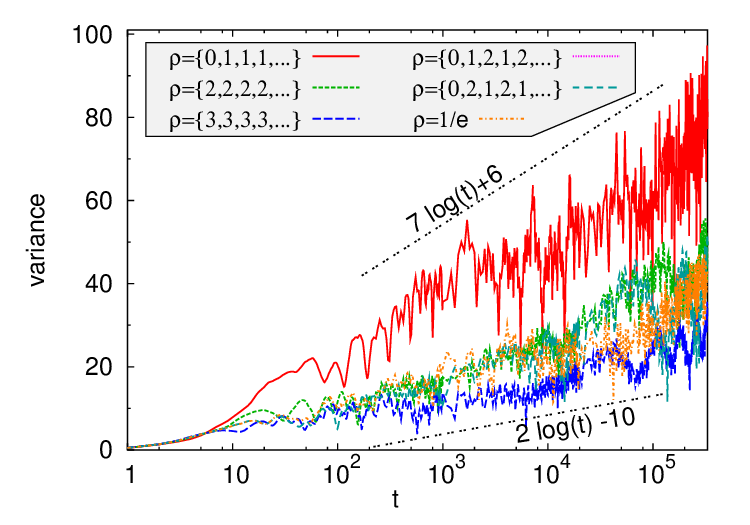}\quad
\includegraphics[width=.35\linewidth]{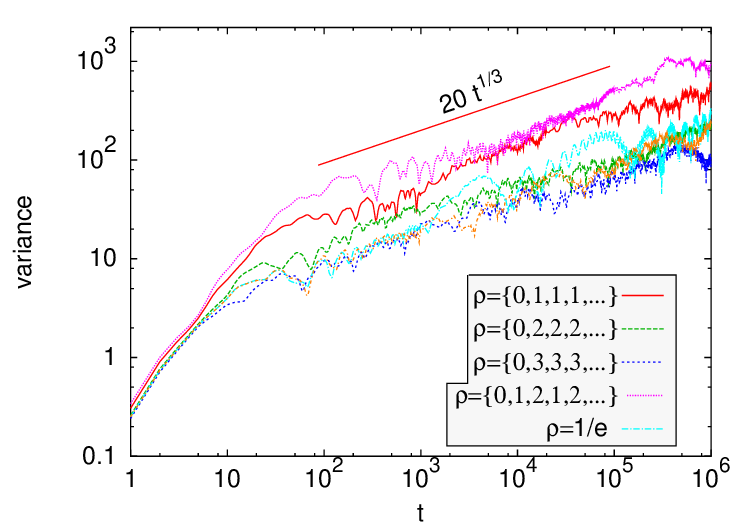}\quad
\includegraphics[width=.35\linewidth]{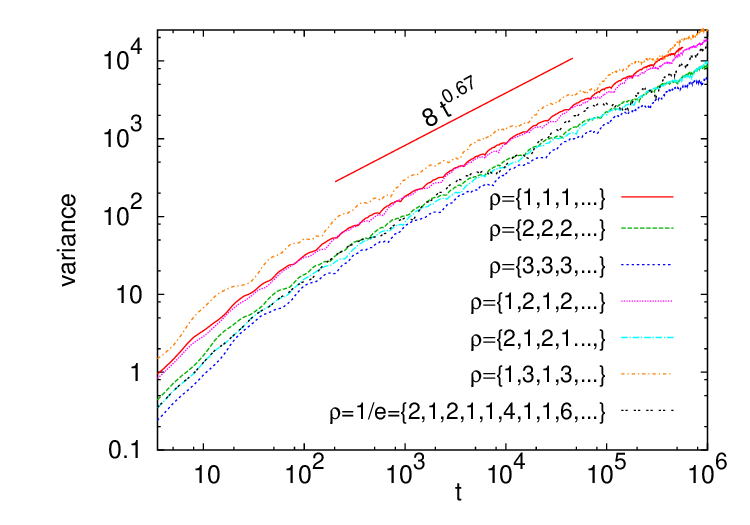}
\caption{Temporal growth of variance for different flow patterns.\\
(a) pattern and parameter values from Fig.~\ref{fig:4patterns}a;\qquad
(b) pattern and parameter values from Fig.~\ref{fig:4patterns}c;\\
(c) pattern and parameter values from Fig.~\ref{fig:4patterns}b;\qquad
(d) pattern and parameter values from Fig.~\ref{fig:4patterns}d.\\
Notation $\{n_0,n_1,n_2,n_3\ldots\}$ indicates first entries in the continued fraction expansion $\rho=n_0+1/(n_1+1/(n_2+1/(n_3+...)))$. Except for $\rho=1/e$, 
all shown continued fractions are periodic.
}
\label{fig:4laws}
\end{figure}

In striking contrast to the conventional case of transport in the presence of molecular diffusion,
variance in the deterministically advected ensemble is not necessarily
monotonically growing with time; during certain epochs, it may decrease.
Although in the course of the quasiperiodic motion the ensemble of non-interacting
tracer particles never exactly returns to any of its previous states, 
it repeatedly comes sufficiently close to all of them so that
the trajectories ``nearly close'', and the system loosely recalls its initial state.
The succession of returns is rigidly encoded in the rotation number,
hence particularities of the temporal behavior of variance depend on the value of $\rho$.
The time intervals between the closest returns are proportional
to the numerators of the best rational approximations to $\rho$.
The latter can be recovered from the expansion of the irrational $\rho$
into the continued fraction
$\rho=n_0+1/(n_1+1/(n_2+1/n_3+\ldots)))\equiv \{n_0,n_1,n_2,n_3,\ldots\}$ by truncating
the expansion after a finite number of entries. Note that $n_0$ is the integer part of $\rho$.
In Fig.~\ref{fig:4laws}a, the dependence var$(t)$ for the flow pattern
without eddies is plotted for three different rotation numbers:
the golden mean $(\sqrt{5}-1)/2=\{0,1,1,1,\ldots\}$, the ``silver mean'' 
$\sqrt{2}-1=\{0,2,2,2,\ldots\}$ and the number $\rho=1/{\rm e}$ with
aperiodic expansion $\{0,2,1,2,1,1,4,\ldots\}$ in the continued fraction. The maximal plotted time 
in Fig.~\ref{fig:4laws}a corresponds to $\sim 10^4$ rotations of a tracer around the torus, 
and the plots bear evidence that in all cases $\mbox{var}(t)$ not only remains bounded,
but stays far below the squared size $(2\pi)^2\approx$39.478 of the single spatial periodicity cell.
This indicates that the ensemble of tracers drifts downstream as a reasonably compact whole.

At $|f|\geq f_{\rm cr}$ the picture is qualitatively different. Two stagnation points (hyperbolic saddles and/or the degenerate points) 
with adjacent eddies are present in the flow pattern. A motion across the vicinity of a saddle
is slow; its duration (``passage time'') diverges when the starting point is shifted closer to the incoming
separatrix of the stagnation point. During such events, the tracer particle in concern lags
far behind the drifting bulk of the ensemble. In the plane, the ensemble gradually gets
stretched along the drift direction. Quantitatively, this is reflected
in the unbounded temporal growth of variance. 
Typical dependencies for the flow pattern with two eddies
are presented in Fig.~\ref{fig:4laws}b
for 6 different irrational values of $\rho$. 
Differing in details, all of them feature a non-monotonic dependence 
on $t$, fluctuating around the logarithmic growth law. 
In the parlance of transport processes, this trend towards
sublinear growth can be referred to as ``subdiffusive behavior''. 

The duration of passage near the hyperbolic stagnation point
is proportional to the logarithm of the initial distance
between the tracer and the incoming separatrix of that point. 
For motions past non-hyperbolic stagnation points
the singularities of passage times are of the power-law nature;
below we show that this, in turn, is reflected by power laws
in temporal evolution of variance: var$(t)\sim t^{\alpha}$. 
As seen in Fig.~\ref{fig:4laws}c, at $|f|=f_{\rm cr}$
when the flow experiences the generic saddle-center bifurcation, 
the growth of variance is remarkably faster:
it fluctuates around the dependence var$(t)\sim t^{1/3}$.
Furthermore, in the inviscid flow pattern shown in Fig.~\ref{fig:4patterns}d
the birth of stagnation points at $|f|=f_{\rm cr}$ is
a degenerate bifurcation; this situation is characterized
by a stronger divergence of the passage time. 
Accordingly, the growth of variance in the ensemble is in this case much faster than in the generic case.
In Fig.~\ref{fig:4laws}d this growth is visualized for 7 different irrational values
of rotation number $\rho$. Although the details differ, in all cases the evolution apparently follows
the power law with the exponent $\approx 0.67$; according to the standard classification, 
this is the \textit{subdiffusive} transport as well.

\subsection{Growth of variance: Trend and decoration}
\label{trend_and_decoration}
\begin{figure}
\centering
\includegraphics[width=.325\linewidth]{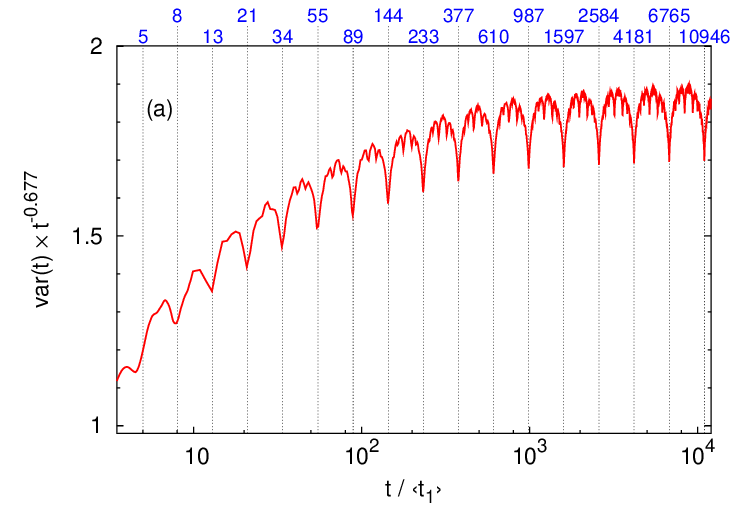}
\includegraphics[width=.325\linewidth]{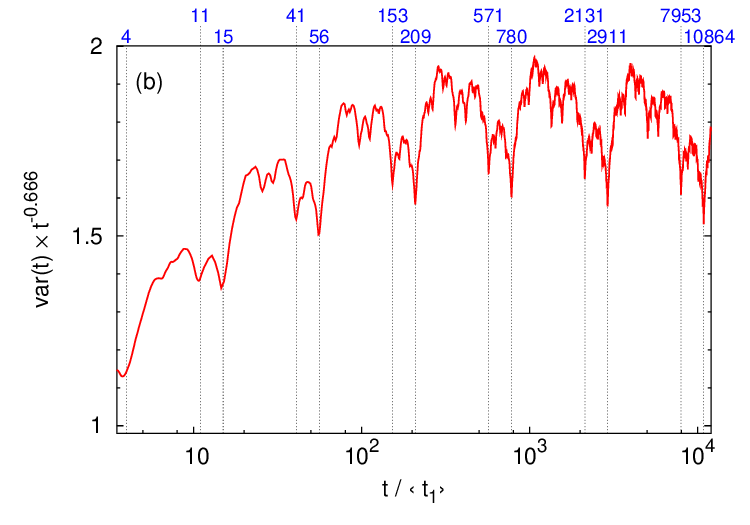}
\includegraphics[width=.325\linewidth]{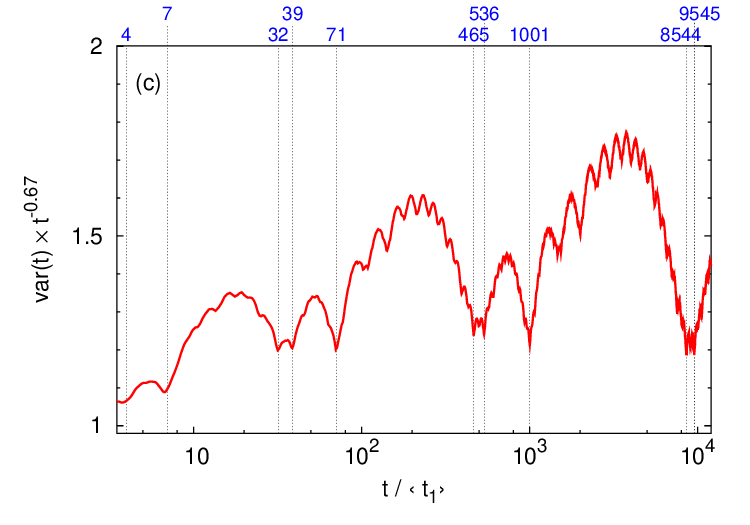}
\caption{``Decoration'' around the power laws $t^{\alpha}$ 
with $\alpha=0.67\pm 0.01$ for the curves from Fig.~\ref{fig:4laws}d.\\
Horizontal axis: time $t$ in units of the average time $\langle T\rangle$ 
for one turn around the torus along the $x$-direction.\\
Vertical dotted lines: numerators of the best rational approximations (see Appendix \ref{app:approx}).\\
(a): the golden mean rotation number $\rho=\left(\sqrt{5}-1\right)/2=\{0,1,1,1,\ldots\}$.\\
(b): rotation number $\rho=\left(\sqrt{3}-1\right)/2=\{0,2,1,2,1,2,1\ldots\}$.\\
(c): rotation number $\rho=1/\mbox{e}=\{0, 2, 1, 2, 1, 1, 4, 1, 1, 6, 1,\ldots\}$.
}
\label{fig:decorations}
\end{figure}

A look at the panels in Fig.~\ref{fig:4laws} shows that neither the logarithmic growth law
nor the power laws are followed closely; rather, they represent some general tendencies,
recognizable at sufficiently large time intervals. 
Recurrent local deviations from these laws follow
a certain pattern, specific to each rotation number.
Functional dependence of the variance on time can be seen
as a combination of two factors: in terms of the phase plane,
one of them is of local origin and is prescribed
by the type of stagnation point, whereas the other one
has to do with the rotation number of the global flow.
The first factor is the  \textit{``trend''},
an unbounded slow monotonic function of time: a logarithmic
dependence (if the stagnation point is hyperbolic) or a power law
if the point is degenerate.
According to our numerical observations, the growth law for the trend
does not significantly depend on the value of irrational rotation
number and quantitatively is largely the same for all stream functions
with the same type of degeneracy of stagnation points.
As discussed above, in contrast to the case of conventional diffusion, the variance of the
deterministically advected ensemble of tracers can temporarily decrease. 
This effect is due to the second factor, which we call below a \textit{``decoration''}.  
Decoration, a fast varying bounded non-monotonic function of time, 
fluctuates on the logarithmic timescale and is prescribed
by the expansion of the rotation number into a continued fraction. 
Local minima of variance arise at the minima of the decoration:
the respective time values of the deepest minima correspond
to the numerators of the best rational approximations
to the rotation number.
If the coefficients of expansion of $\rho$ into the continued 
fraction build the 
periodic sequence, i.e. $\rho$ is a quadratic irrational,
then the time values corresponding to the deepest minima of decoration
asymptotically form the log-periodic sequence.
In the case of a generic $\rho$, the deepest minima of decoration
are separated by irregular time intervals. Fig.~\ref{fig:decorations} 
illustrates the decoration patterns (that is, of variance divided
by the respective trend) for three exemplary irrational
rotation numbers: (a) the golden mean, 
(b) the number with expansion into continued fraction of period 2, 
and (c) the number with aperiodic continued fraction.

\subsubsection*{Self-similarity in the decoration pattern}
\begin{figure}[h]
\centering
\includegraphics[width=.99\linewidth]{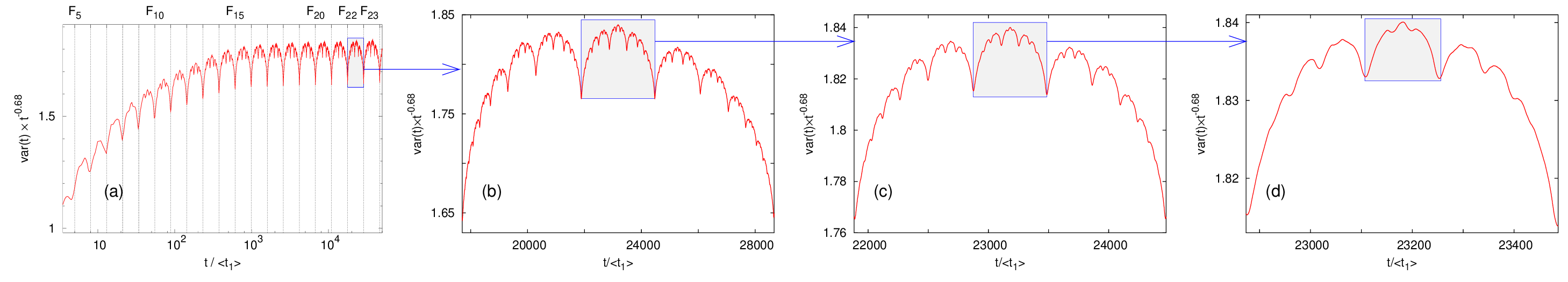}
\caption{Self-similarity of decoration for the curve from Fig. ~\ref{fig:decorations}a:
growth of variance in the ensemble of particles carried by the flow~(\ref{velocity_field}) 
with the golden mean rotation number $\sigma$.
Parameter values: $\alpha=(\sqrt{5}-1)/2,\,\beta=1,\,\nu=0,\,f=f_{\rm cr}=0.61803$.\\
Horizontal axis: time $t$ in units of the average time $\langle T\rangle$ 
for one turn around the torus along the $x$-direction.\\
Vertical dotted lines: Fibonacci numbers $F_n$.\\
(a) the  overall diagram for $t/\langle T\rangle \leq F_{24}$=46368.
(b) magnified rectangular area from (a): interval between $F_{22}$ and $F_{23}$.\\
(c) magnified rectangular area from (b); (d) magnified rectangular area from (c).
}
\label{fig:similarity}
\end{figure}

The left panel of Fig. \ref{fig:decorations} may leave the (false) impression that
for the values of $\rho$ with periodic expansion into the continued
fraction, the profile of decoration is just log-periodic with respect to time.
However, the picture is far more involved.
The deepest minima of decoration indeed asymptotically form the log-periodic 
lattice, but while moving along this lattice in the direction of growing time $t$,  
the inter-minima intervals become more and more fragmented, 
with formation of new minima and maxima. 
Patterns in the segments between the deepest  \textit{secondary} minima bear,
on a smaller scale, resemblance to the patterns between the deepest minima,
featuring the sublattices of sharp and relatively deep \textit{tertiary} minima etc.
This is, in a sense, a kind of self-similarity,
albeit in the reverse direction: from small through moderate to ever larger scales. 
In contrast to conventional geometric fractals like the Koch curve, 
here the self-similar picture is unfolding not in space but in time, 
and in order to properly capture it, we need, instead of the infinitely fine spatial resolution, 
the infinitely long time interval. We illustrate this phenomenon in Fig.~\ref {fig:similarity}
for the case of the golden mean rotation number $\sigma$.
Here, the temporal pattern of variance in the dimensionless time units
is organized around the lattice of Fibonacci numbers.  
Scaling factors are dictated by the rotation number:
Subplots (c) and (d) are obtained from preceding subplots, respectively, (b) and (c),
by magnification of respective central parts with the factor
$\sigma^{-3}=2+\sqrt{5}$ in the horizontal direction
and $\sigma^{-2}=(3+\sqrt{5})/2$ in the vertical direction.

\section{Mathematical models}
\subsection{From flows of fluids to plane Hamiltonian flows}
\label{sect:passage_time}
In the previous section, we have presented evidence that differences
in the temporal growth of variance for ensembles of drifting tracer
particles are related to the slowdowns that the particles
experience while moving past the stagnation points of the velocity field. 
In this section, we determine how long it takes for an advected particle
to traverse the region of the stagnation point. 
Starting with the generic case of the hyperbolic stagnation point,
we proceed to degenerate situations. 
\blue{Transformations of steady flow patterns in the course of
bifurcations of stagnation points are illustrated e.g. in~\cite{Bakker_book}. 
Here, we focus on the impact of degeneracies on the
distributions of passage times across the relevant flow regions.}

Let the Cartesian coordinates on the plane be denoted by $x$ and $y$
(which are not necessarily the original coordinates of the hydrodynamical problem (\ref{Navier_psi})). 
For the sake of simplicity, we restrict ourselves to the class of flow patterns
generalizing Eq.~(\ref{stat}) in the following sense:
in the chosen coordinates the representation of the stream function is additive
(in fact, we require less: that the leading terms of its Taylor expansion in $x$ and $y$ are monomials): 
$\Psi(x,y)=\psi_1(x)+\psi_2(y)$.\footnote{In the related Hamiltonian description
this corresponds to the conventional decomposition into the 
coordinate-dependent potential energy and momentum-dependent kinetic energy; 
in problems of mechanical origin the latter dependence is quadratic.}
Then, the velocity along each direction is the function of the complementary coordinate alone: 
$v_x=v_x(y)=\displaystyle \frac{d\psi_1(y)}{dy}$ and 
$\displaystyle v_y=v_y(x)=-\frac{d\psi_2(x)}{dx}$.
We shift the stagnation point to the origin $x=y=0$
and expand both velocity components near this point into the Taylor series.
Let positive integers $m$ and $n$ denote the exponents of the first non-vanishing terms in the Taylor expansions of $v_x$ and $v_y$:
$v_x(y)=a_m y^m+\mathcal{O}(|y|^{m+1})$
and $v_y(x)=b_n x^n+\mathcal{O}(|x|^{n+1})$ with nonzero $a_m$ and $b_n$.

The case $m=1$ includes conventional conservative systems
of mechanical origin like $\ddot{x}=f(x)$, 
where equations on the plane with the state of equilibrium
at $x=y=0$ can be brought to the form : 
\begin{eqnarray}
\label{dx_1}
\dot{x}&=&y\\
\label{dy}
\dot{y}&=&f(x)=({n+1})\,x^n+o(|x^n|)
\end{eqnarray}
with $n=1$ for the structurally stable \textit{non-degenerate} hyperbolic state of equilibrium
and $n\geq 2$ in the case of various degeneracies. 
            
\begin{figure}[h]
\centering
\includegraphics[width=.7\linewidth]{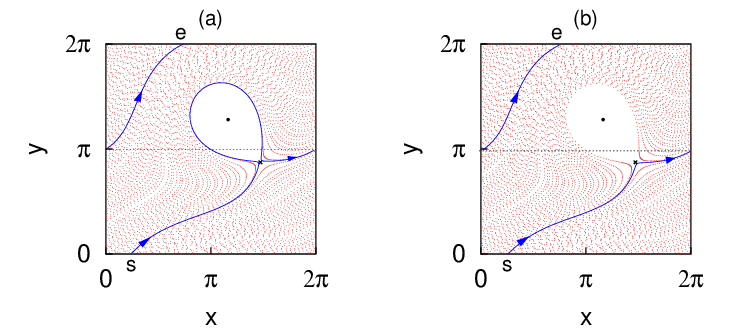}\quad
\caption{\blue{Geometric reason for asymmetric distribution of passage
times near the hyperbolic stagnation point.\\
Crosses: hyperbolic stagnation points. Filled circles: elliptic stagnation points.\\
Dotted horizontal line: ``gluing'' of trajectory segments while intersecting the vertical torus ``borders''.\\
Panel (a): passage with two slowdowns along the separatrix, encircling the vortex. Passage time from $y=0$ to $y=2\pi$: 22.83.\\ 
Panel (b): direct passage across the cell with a single slowdown. 
Passage time from $y=0$ to $y=2\pi$: 12.21.\\
Symbols s and e denote, respectively, starting points 
of trajectories at $y=0$ and their endpoints at $y=2\pi$.} 
}
\label{fig:asym_passage}
\end{figure}

The system (\ref{dx_1}-\ref{dy}) has the obvious integral 
$H=-x^{n+1}+\displaystyle\frac{y^2}{2}$.
Separatrices $y=\pm \sqrt{2}\,x^{(n+1)/2}$ of the saddle point at the origin correspond to $H=0$; there are four separatrices for odd $n$ and two separatrices 
for even $n$.
For $H<0$
the tracers, governed by the equation 
\begin{equation}
\dot{y}=(n+1)\left(\frac{y^2}{2}-H\right)^{\displaystyle\frac{n}{n+1}} 
\end{equation}
approach the saddle along the separatrix with negative $y$
and depart from it along the separatrix with positive $y$.
\blue{On the streamline defined by $H$, we choose two 
fixed values: $y_A<0$ and $y_B>0$.
The time of passage between $y_A$ and $y_B$ equals
\begin{equation}
T_{AB}(H)=\frac{1}{n+1}\int_{y_A}^{y_B} \frac{dy}{\displaystyle\left(\frac{y^2}{2}-H\right)^{\displaystyle\frac{n}{n+1}}}. 
\label{T_AB}
\end{equation}

We begin the analysis of (\ref{T_AB}) with the hyperbolic case $n=1$.
Here, the integration  yields
\begin{equation}
\label{passage_time_hyperbolic}
T_{AB}(H)=\frac{1}{\sqrt{2}}\,\log\displaystyle
\frac{y_B+\sqrt{y_B^2-2H}}{y_A+\sqrt{y_A^2-2H}}.
\end{equation}
At small values of $|H|$ (recall that both $H$ and $y_A$ 
are negative), the latter expression tends to
$\displaystyle\frac{1}{\sqrt{2}}\,\log \frac{2y_A y_B}{H}$;
in the limit $H\to 0$, the passage time diverges 
as $2^{-1/2}\log(1/|H|)$.

Let $x_{\rm en}$ denote the coordinate value at the entrance
to the periodicity cell for the streamline carrying the test particle. 
Further, let $s_{\rm en}$ be the analogous coordinate value
for the incoming separatrix of the stagnation point.
 
On a streamline close to the separatrix,  
$H\sim -|x_{\rm en}-s_{\rm en}|$, 
and the duration of the passage 
through the whole periodicity cell 
turns into
\begin{equation}
\label{T_log}
T_{\rm passage} \approx A_{1,2} \log\frac{1}{|x_{\rm en}-s_{\rm en}|}
\end{equation}
Remarkably, the prefactors $A_i,\;i=1,2$ \  
for the passages on two opposite sides of the saddle
(e.g. ``left from the incoming separatrix'' and ``right from 
the incoming separatrix'') differ. This difference is a consequence of the peculiarity of
conservative systems: in them, one of the outgoing
separatrices of the hyperbolic saddle point generically returns back
to the saddle, forming the homoclinic loop around the adjacent
vortex centered at the elliptic stagnation point and filled
by closed phase curves. 
While moving along this separatrix, a tracer experiences
the slowdown twice: first, during the original approach 
to the saddle,  and then at the return to the immediate 
vicinity of the saddle. 
In contrast, a tracer moving along the complementary outgoing
separatrix leaves the vicinity of the saddle for good
and does not return; it experiences the slowdown only once, 
hence its passage duration is a half of the overall passage
duration of a tracer that closely follows the vortex boundary. 
This means that the passage time as a function of the entrance coordinate possesses the \textit{asymmetric} logarithmic singularity; the prefactors differ by a factor 2.
The configuration of
streamlines which causes this asymmetry in the passage durations
is sketched in Fig.~\ref{fig:asym_passage}}.

Let us proceed to the degenerate flows with $n>1$.
For $y_A<0<y_B$, at small $|H|$ the integral (\ref{T_AB}) is
dominated by the contribution of the segment near $|y|=0$
and is well approximated by its value, obtained in the limit
$y_A\to -\infty,\,y_B\to\infty$:
\begin{equation}
T_{\rm lim}(H)=C H^{-\kappa}
\blue{\approx C|x_{\rm en}-s_{\rm en}|^{-\kappa}}
\end{equation}
where 
\begin{equation}
\label{kappa_old}
\kappa=\,\frac{n-1}{2(1+n)}=\frac{1}{2}-\frac{1}{n+1}\,
\end{equation}
and the (non-important) constant $C$ 
is
$$C=\frac{\sqrt{2\pi}\,\Gamma(\kappa)}{\displaystyle\Gamma\left(\frac{n}{n+1}\right)}.$$
Since at large $|y_A|$ the distance between the arbitrary 
\blue{streamline and the separatrix
is proportional to the value of $H$ on this streamline, we obtain
the power law singularity
\begin{equation}
\label{T_passage}
T_{\rm passage}\sim |x_{\rm en}-s_{\rm en}|^{-\kappa}.
\end{equation}
}
\medskip

In a wider, non-mechanical context certain higher degeneracies cannot be brought to the form (\ref{dx_1}-\ref{dy}): 
Eq.(\ref{dx_1}) should be replaced by
\begin{equation}\dot{x}=y^m\;\;\Rightarrow\;\ddot{x}\sim {\dot{x}}^{m-1}x^n,
\label{replacement}
\end{equation}
with $m\geq 1$. The time of passage near the equilibrium obeys the same law (\ref{T_passage}) where 
Eq. (\ref{kappa_old}) is generalized to 
\begin{equation}
\label{kappa_new}
\kappa=\frac{m\,n-1}{(m+1)\,(n+1)}=\frac{m}{m+1}-\frac{1}{n+1}.
\end{equation}
The value of $\kappa$ is below 1, but unlike  (\ref{kappa_old}), 
Eq.(\ref{kappa_new}) renders in certain cases (e.g. $m=n>3$) the values 
exceeding $1/2$; as we shall see below, this results in superdiffusive transport for the corresponding fluid motion.

For the velocity field (\ref{velocity_field})
only a few combinations of $m$ and $n$ are relevant.
The case $m=n=1$ describes the non-degenerate (hyperbolic) stagnation point. 
The case $m=2,\,n=1$ (or equivalently $m=1,\,n=2$)
corresponds to the saddle-center bifurcation in a conservative system: birth of a structurally unstable point
of equilibrium which, under the parameter variation, splits into the elliptic point and the hyperbolic point. 
The case $m=n=2$ occurs in the family (\ref{velocity_field}) only in the inviscid fluid
with $\nu=0$. In order to feature degeneracies of higher orders, the stream function should include
Fourier harmonics that have higher order in $x$ and $y$; this can be achieved by including the appropriate higher harmonics in
the spatial pattern of the force ${\bm F}$ 
in equations (\ref{navier}) and (\ref{Navier_psi}).

\blue{
\subsection{Further relevant flow patterns}
Here we characterize two further types of passage times
singularities that can be encountered in plane flows but do not 
show up in the frame of the family of velocity fields (\ref{velocity_field}).  

As mentioned above and illustrated in Fig.~\ref{fig:4patterns}(c), the eddies in the flow pattern (\ref{stat}) 
exist pairwise; they are counterrotating and are born in 
saddle-center bifurcations that simultaneously happen in 
two different points of the torus at the value $f_{\rm cr}$ 
of the forcing amplitude. 
Accordingly, at $f>f_{\rm cr}$ there are two hyperbolic 
stagnation points inside each periodicity cell, each one 
with its own adjacent eddy and two incoming separatrices. 
Hence an adequate special flow for this configuration 
requires the time function $f(u)$ with \textit{two} 
asymmetric logarithmic singularities.
Remarkably, the adjacent eddies lie on the opposite sides of
stagnation points, therefore the singularities in $f(u)$ 
effectively balance each other:
the sum of both prefactors before the logarithmic terms 
on the left from the singularities is practically equal 
to the sum of both prefactors on the right. As discussed above,
this circumstance hampers mixing~\cite{Kochergin-2007}.
A different situation would occur in a flow pattern with just
one hyperbolic stagnation point (and hence a single eddy) 
on the surface of the torus: there, $f(u)$ is generically 
non-balanced, and the faster temporal growth of variance 
in the ensemble of tracers is expected. Absent in the frame 
of the family (\ref{stat}), such flow patterns can be excited
by sufficiently strong spatially periodic forcing 
with more elaborate geometry.
For example, a flow pattern with a single saddle point per periodicity cell occurs in the family of stream functions
\begin{equation}
\label{psi_one_vortex}
\Psi(x,y)=\alpha  y - \beta  x + q\, (\sin x \cos y - \sin x - \cos y)
\end{equation}
Here, the linear part is the same as in (\ref{velocity_field}),
whereas the terms, periodic with respect to spatial coordinates,
are different. Like in (\ref{velocity_field}),
the rotation number is $\alpha/\beta$. Presence/absence of
stagnation points depends on the value of the parameter $q$:
below the threshold $q_{\rm thr}$ the linear terms dominate and the velocity field has no singular points. At the threshold
$q=q_{\rm thr}$ the saddle-center bifurcation takes 
place\footnote{$q_{\rm thr}=\displaystyle
\frac{6\sqrt{6}\,\alpha\beta}
{\sqrt{(\alpha^2+\beta^2)(34\alpha^2\beta^2-\alpha^4-\beta^4)
+(\alpha^4+14\alpha^2\beta^2+\beta^4)^{3/2}}}$}, 
creating the single degenerate stagnation point which, at the
increase of $q$, splits into the saddle and the elliptic point; the latter is located inside the eddy filled by closed streamlines and 
adjacent to the hyperbolic point. The corresponding
flow pattern is visualised in Fig.~\ref{fig:asym_passage}.

\medskip

Another important class of fluid motions with 
singularities of passage time comprises steady plane flows 
of viscous fluids through periodic arrays of solid obstacles. 
In the approximation of the Stokes flow (creeping flow), 
motions of this kind are solutions of the biharmonic equation
for the stream function: $\Delta\Delta\Psi=0$. 
A characteristic pattern of the flow through the square lattice 
of circular obstacles is visualised in Fig.~\ref{fig:stokes_flow}(a).
Hydromechanical properties of such viscous flows 
were studied in~\cite{Hasimoto-59,Sangani-Yao}.
Here, like in the above cases, the stream function is a superposition
of doubly periodic component with the component, linear in both
coordinates, therefore the elementary periodicity cell is 
tantamount to the torus surface.
We note, however, a crucial distinction: in this case, 
on the torus there is an impenetrable material border, 
the boundary of the obstacle. 
In presence of viscosity, the boundary conditions require
that both components of velocity vanish
along the whole boundary of each obstacle.
In a sense, the boundary can be viewed as a one-dimensional
continuum of stagnation points. As the flow drags the tracer
particles along the continuum, the collective action of the
multitude of stagnation points generates the strong singularity
of the passage duration. If the rotation number on the torus
-- the mean inclination of streamlines to the coordinate axes --
is irrational, the Lagrangian observables (e.g., the velocity
components of a tracer particle, drifting in this velocity field)
possess fractal Fourier spectra whereas the envelope of the 
autocorrelation of velocity decays in accordance with a power 
law~\cite{stokes_PRL,stokes_PTP}; transport of tracers 
by such flows was found to be weakly superdiffusive.

\begin{figure}
\centering
\includegraphics[width=.2\linewidth]{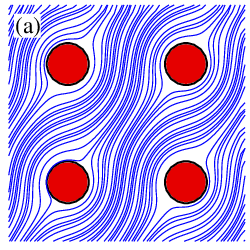}\qquad\qquad
\includegraphics[width=.18\linewidth]{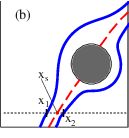}
\caption{\blue{Stokes flow of a viscous fluid past a doubly periodic array of solid obstacles.\\
(a) Pattern of streamlines for a flow with irrational rotation number.
(b) Motion past an obstacle.
Solid curves: two exemplary streamlines. 
Dashed line: separatrix between passages beyond the obstacle 
and passages beneath the obstacle.
Dotted line: a secant at $y$=const, intersecting, respectively, 
the separatrix at $x_s$ and the streamlines at $x_{1,2}$.
Duration of the passages $~\sim |x_i-x_s|^{-1/2},\;i=1,2$.}}
\label{fig:stokes_flow}
\end{figure}

The reason for anomalies is the unbounded slowness of motions
in the layer adjacent to the solid border.
While circumventing an obstacle, the stream splits and
reassembles again; as seen in Fig.~\ref{fig:stokes_flow}(b), 
for each obstacle there is a streamline that begins/ends on its
border and separates the streamlines surpassing the obstacle
from the opposite sides. The time of passage diverges 
while approaching this separatrix; it can be shown 
that the duration of passage along the
particular streamline is inversely proportional to the
square root of the distance at the entrance of the cell 
between this streamline and the incoming separatrix of the obstacle:
\begin{equation}
\label{square_root_law}
T(x_{\rm en})\sim\frac{1}{\sqrt{|x_{\rm en}-s_{\rm en}|}}
\end{equation}
with $x_{\rm en}$ and $s_{\rm en}$ denoting the entrance coordinates of, respectively, the streamline in question
and the separatrix.

Formally, the relation (\ref{square_root_law}) describes the same 
strength of singularity as the relation (\ref{T_passage}) at 
$\kappa=1/2$. In accordance with Eq.(\ref{kappa_new}), this
matches the already mentioned high-order singularities in 
Eq.~(\ref{replacement}) with $m=n=3$ and ($m=2,\;n=5$), or vice versa. 
However, in contrast to the high-order singularities in (\ref{replacement}) that 
are vulnerable with respect to generic deformations of the velocity field, 
the robust geometry of the flow pattern in Fig.~\ref{fig:stokes_flow}(a)
makes the inverse square root singularity of passage times
structurally stable: given the array of obstacles, 
minor changes of the velocity field may alter
the rotation number, but not the dependence (\ref{square_root_law}).
}

\subsection{Modeling by special flow}
\label{subsect:special_flow}
In a steady plane flow, the tracer particles drift
along the streamlines.
Every streamline belonging to the global component of the flow carries the tracers 
across the infinite row of elementary periodicity cells. 
It is natural to discretize the motion by considering
a kind of Poincar{\'e} mapping induced by the flow. 
We choose for this purpose the one-dimensional mapping $x_j\mapsto x_{j+1}$ which interrelates
the values of the coordinate $x$ of subsequent intersections of the streamline
with the borders of the elementary cells: 
lines $y=2\pi j,\,j=0,\pm1,\ldots$. 
For example, for every panel in Fig.~\ref{fig:4patterns}, 
the mapping interrelates the values of the horizontal coordinate $x$
on the horizontal borders of elementary cells.
By construction, the mapping is a monotonic (streamlines cannot intersect!) circle map. 

Importantly, transport characteristics cannot be recovered
from the trajectories of the mapping alone. 
The reason: These characteristics are time averages
obtained in the limit of $t\to\infty$, whereas
the mapping procedure ignores the fact that a transported tracer
spends time intervals of different lengths in different cells. 
In the presence of saddle stagnation points, 
the duration of the passage across the cell as a function
of the entry position is unbounded. 
A passage relatively close to the stagnation point takes longer
and makes a bigger contribution into the average characteristics 
(e.g. variance) than the relatively fast passage further from stagnation points. 
However, in conventional mappings all iterations share the same weight. 
To obtain an adequate description for the flow, the
duration of the passage should be taken into account:
this would allow us to assign higher weights to long
slow passages from the elementary cell to the next elementary cell,
and, respectively, lower weights to short fast passages.
A modification of the mapping that assigns ``duration'' 
to its iterations is known
in ergodic theory as the ``special flow''~\cite{CFS_book},
and goes back to von Neumann~\cite{von_Neumann}.

\begin{figure}
\centering
\includegraphics[width=.5\linewidth]{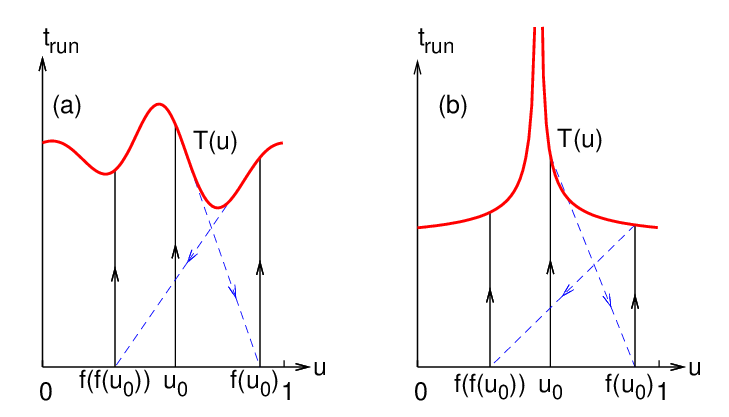}
\caption{Construction of the special flow over the circle map.
Coordinate $u$ is the mapping variable; coordinate $t$ corresponds
to time.
Red curve: ``function over the mapping'' $T(u)$ which mimics the
time of passage through the elementary cell of the flow along the
streamline that enters the cell at coordinate value $u$.
Black vertical lines: segments of phase trajectory.
Blue dashed lines: instantaneous jumps of $u$ corresponding
to subsequent iterations of the circle map. $u_0$: starting value of $u$ for iterations of the circle map.\\
Panel (a): continuous bounded $T(u)$: a flow
with bounded passage time.
Panel (b): unbounded $T(u)$, corresponding to
a flow past the saddle stagnation point.
}
\label{fig:specflow}
\end{figure}

A mapping variable, denoted below as $u$, is the
properly rescaled (say, so that $0\leq u <1$) monotonic function
of the original variable, in our case of $x$.
Given the mapping $u\mapsto f(u)$, the special flow is built over it
in the following way (see, e.g., Chapter 11 of~\cite{CFS_book}): 
As sketched in Fig.~\ref{fig:specflow}, 
the coordinates in the phase space are
the variable $u$ of the mapping and \blue{the running time $t_{\rm run}$} 
of the current iteration of the mapping; 
in our setup, the latter is the time, elapsed since the entrance 
into the current elementary cell. The phase space of the special flow is
the plane region between the abscissa 
and the prescribed function $T(u)$. The latter corresponds
to the full passage time through the elementary flow cell of the
tracer that enters the cell at the coordinate value $u$.
The phase trajectory of the special flow
consists of discontinuous vertical segments
that connect the abscissa to the line $T(u)$.
Dynamics evolves as follows: the variable $u$ is piecewise constant,´
whereas dependence of the variable $t_{\rm run}$ on the time is 
linear with unit slope and piecewise continuous. 
Evolution starts at some point $(u_0,0)$. The value of $u$ stays
fixed whereas the value of $t_{\rm run}$ increases 
with unit speed $(\dot{t}_{\blue{\rm run}}= 1)$
until reaching the value $T(u_0)$; from there, the system
instantaneously jumps into the position 
$(f(u_0) \rm{mod\, 1},\,0)$,
and begins the next epoch of upward vertical motion with unit speed. This motion goes on until 
hitting the curve $T(u)$ again, then the next jump to abscissa follows, and so on. 
Von Neumann demonstrated that the special flow, built over the mapping
with purely discrete spectrum and possessing discontinuities in $T(u)$,
can contain the continuous spectral component~\cite{von_Neumann}. 
Later, special flows with different kinds of singularities 
were used for studies of spectral and mixing properties
of the flows on 2-tori.
Kolmogorov~\cite{Kolmogorov-53} showed that in the absence
of points of equilibrium on the surface of the torus
the spectrum was discrete for a broad class of irrational 
rotation numbers
\textit{(Diophantine} numbers, not too fast approximated by rationals),
and could be continuous for other irrationals
(see also~\cite{Shklover-67}). In the presence of degenerate points
of equilibrium (that is, with power-law singularities
in the function $T(u)$ of the special flow), the mixing property for all irrational rotation numbers was established in~\cite{Kochergin-75}. 
For non-degenerate (hyperbolic) saddle points, the singularities of $T(u)$
are logarithmic. 
If the sum over all saddles of all prefactors at the logarithmic terms
on the left from the respective saddles coincides with the analogous
sum of all prefactors on the right, $T(u)$ is called a symmetric (balanced) function with logarithmic singularities. 
Kochergin established that in the case of a symmetric function, 
the special flow is non-mixing~\cite{Kochergin-76,Kochergin-2007}.  
For the case where the symmetry is absent, Sinai and Khanin showed existence of mixing~\cite{Sinai-Khanin-92}. 
Notably, flow patterns on a torus with a single saddle stagnation point
(in terms of a lift onto a plane, these are the flows
with a single hyperbolic point per periodicity cell) cannot be
adequately represented by a special flow with a symmetric function:
\blue{as discussed above, the prefactor on one side of the logarithmic singularity} is
twice as large as the prefactor on the other side, hence such
flows always mix, albeit \blue{the mixing processes in them 
run slower} than in the flows with degenerate stagnation points.
 
\blue{A differentiable circle map with positive derivative 
that has bounded variation is, under an irrational value of rotation number $\rho$,} conjugate to the linear shift 
$u\mapsto (u+\rho)\mod 1$. Later on, we choose $u=u(x)$ 
in such a way that $f(u)=(u+\rho)\mod 1$.

\blue{To assess the transport properties of the special flow,
we introduce the \textit{phase} $\varphi(t)$ of the flow: 
at the time $t$,  the integer part of $\varphi$
is the number $n(t)$ of already accomplished iterations of the circle map 
whereas the fractional part of $\varphi$
measures the relative position of the system on its way upwards 
on the phase plane, reset to 0 at the beginning of each iteration 
and uniformly growing to $1$: 
$$\varphi(t)=n\,+\,\frac{t_{\rm run}(t)}{T(u_n)}.$$
By construction, $\varphi(t)$ is monotonic and continuous. 
The evolution of the central moments of $\varphi(t)$
characterizes the transport.

\section{Special flow: Results of numerical simulations}
\label{sect:numeric}
In the dynamical system that governs the motion of a passive tracer
carried by the plane steady incompressible flow, both Lyapunov
exponents vanish. In the absence of positive Lyapunov exponents, the
mixing processes are slow. Inhomogeneities of the velocity field
contribute to mixing, but convergence of variance (\ref{variance}),
as well as of the higher moments of the distribution, occurs slowly,
especially in the case of relatively weak logarithmic singularities
of the passage time.
Therefore, accurate determination of transport characteristics
requires numerous passages across elementary cells of the flow.
This makes numerical estimates through direct simulations
of the hydrodynamical equations rather time-consuming.
Furthermore, in the course of numerical integration of the
Navier-Stokes equations over very long time intervals, 
it is difficult to avoid accumulation of errors.
In this sense, special flows serve for remarkable improvement
in terms of both the speed (instead of integration of PDEs
or ODEs, only iterations of the circle shift are performed)
and the accuracy (for an iteration of the explicitly defined map,
the error is essentially the processor round-off error).

\subsection{Growth laws of variance for flows with exemplary singularities}

Formally, transport properties of special flow can be studied
for arbitrary singularities of the passage time $T(u)$. 
For an illustration, we take the exemplary singularities which correspond to typical cases arising in fluid motions past the stagnation points:

a) A flow pattern with a single hyperbolic saddle point per periodicity cell
is represented by the special flow in which $T(u)$ has an asymmetric logarithmic singularity,
with the prefactors on two sides of the singularity
differing by the factor $2$, see Eq.(\ref{T_log}).

b) A special flow with the weakest relevant power law singularity of $T(u)$
models the saddle-center bifurcation in the pattern of streamlines;
the corresponding singularity of $T(u)$ is $(u-u_0)^{-1/6}$.

(c) In a special flow mimicking the pattern of motion of the viscous fluid
past the regular array of solid obstacles, $T(u)$ has a singularity
of the kind $(u-u_0)^{-1/2}$.

In all three cases we use for the mapping the circle shift with the
golden mean rotation number.

\begin{figure}[h]
\centering
\includegraphics[width=.3\linewidth]{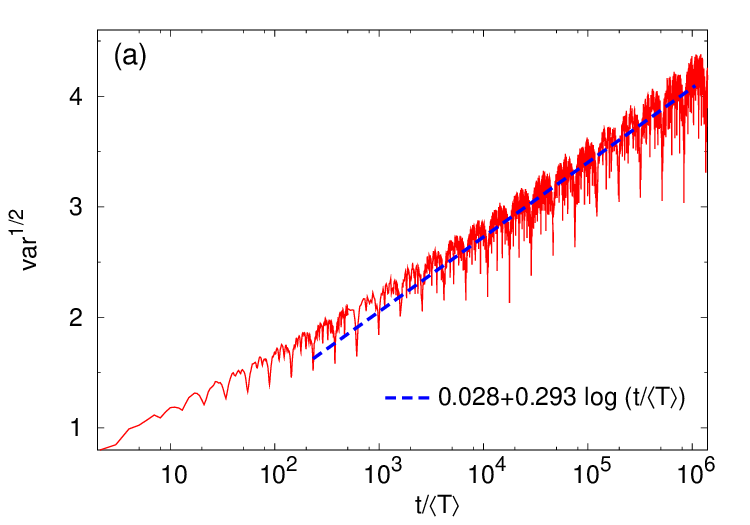}\quad
\includegraphics[width=.3\linewidth]{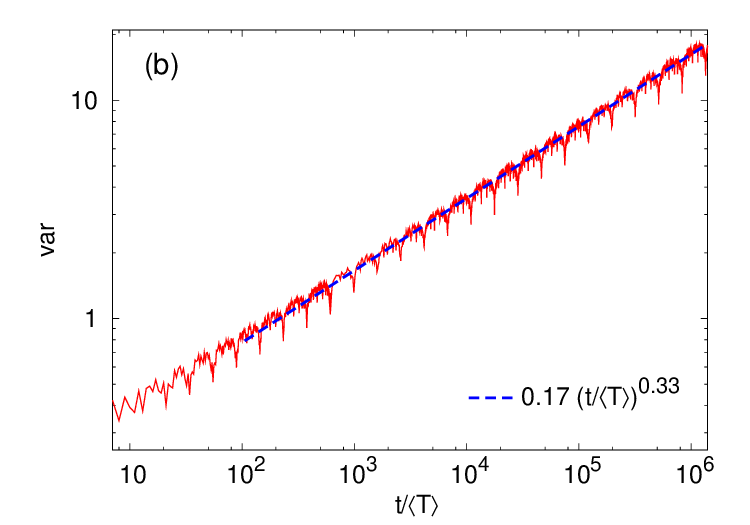}\quad
\includegraphics[width=.3\linewidth]{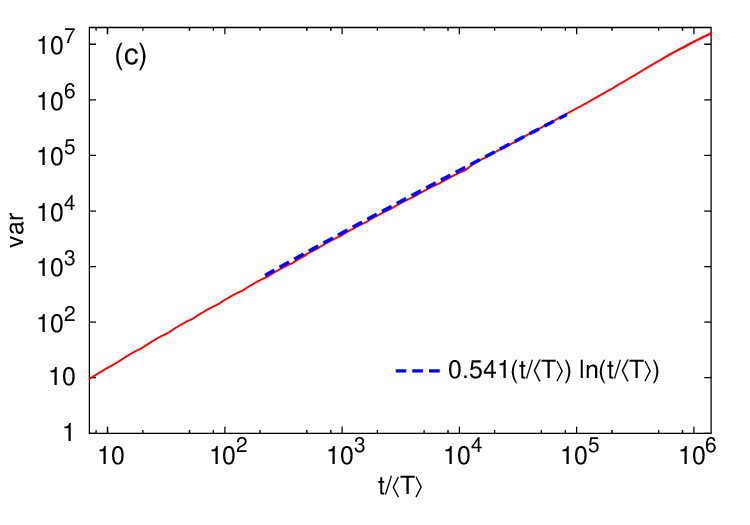}\quad
\caption{Time dependence of variance in special flows 
over the circle map on the interval $[0,1)$\\
with golden mean rotation number $\rho=(\sqrt{5}-1)/2$.\\
Time is expressed in units of average duration $\langle T\rangle$
of one iteration of the circle map.\\
Dashed lines: fitted dependencies.\\
(a) asymmetric logarithmic singularity of passage time:
$T(u)=\left\{\begin{array}{lc} 1-\ln(1/2-u)|,& u\le 1/2\\
1-\ln 2 -2\ln(u-1/2),& u>1/2\end{array}\right.,\;
\langle T\rangle=\displaystyle\frac{5}{2}+\ln 2$;\\
(b) weak power-law singularity corresponding to the 
saddle-center bifurcation: $T(u)=2\,|u-1/2|^{-1/6},\;
\displaystyle\langle T\rangle=\frac{12}{5}\,2^{1/6}$;\\
(c) special flow with strong power-law singularity: 
$T(u)=2|u-1/2|^{-1/2},\;\langle T\rangle=2\sqrt{2}$.
}
\label{fig:specflow_tran}
\end{figure}
We observe that like in the simulations of continuous equations 
from Sect.~ \ref{ref:sect_hydr}, the temporal growth of variance 
can be decomposed into the trend and the decoration. 
Quantitatively, the trend in the case of Fig.~\ref{fig:specflow_tran}(a) 
goes approximately as $\ln^2(t)$, corresponding to weak subdiffusion. 
The trend of variance in Fig.~\ref{fig:specflow_tran}(b) shows the faster
subdiffusion with proportionality to $t^{1/3}$. Finally, the case of
strong singularity in ~\ref{fig:specflow_tran}(c) provides an example of
weakly \textit{super}diffusive transport with ${\rm var}(t)\sim t\ln t$.
Decorations, with roughly log-periodic pattern and characteristic minima
at the Fibonacci numbers, are well recognizable in the left and central panels
of Fig.~\ref{fig:specflow_tran}.
In the right panel, the decoration is masked by the remarkably rapid growth 
of variance; to resolve it, we divide the computed ${\rm var}(t)$ by the trend 
$t\ln t$ (Fig.~\ref{fig:specflow_decor}). 

\begin{figure}[h]
\centering
\includegraphics[width=.3\linewidth]{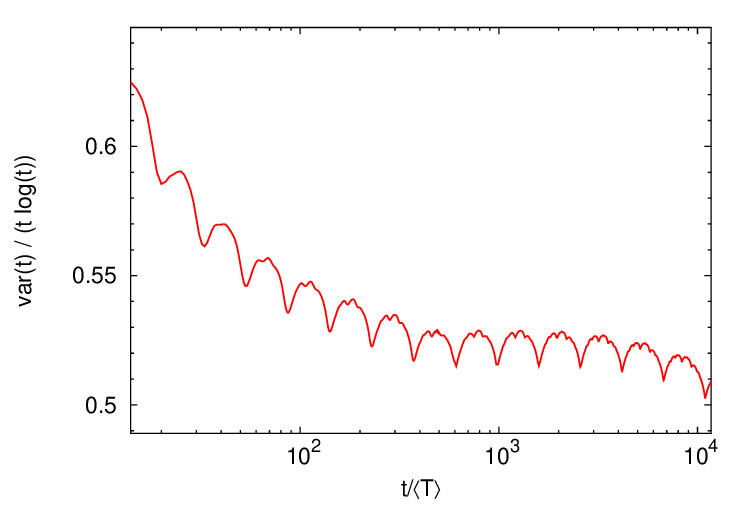}\quad
\caption{Decoration pattern corresponding to the panel (c)
of Fig.~\ref{fig:specflow_tran}. Special flow 
with the strong singularity of $T(x)$ over the circle map
with golden mean rotation number $\rho=(\sqrt{5}-1)/2$.
$T(u)=2|u-1/2|^{-1/2},\;\langle T\rangle=2\sqrt{2}$.}
\label{fig:specflow_decor}
\end{figure}

}



\section{Theoretical predictions}
\label{sect:theory}
In the present section, we develop simplified models of \green{the particle dispersion}
based on the analysis of special flows.

Consider a special flow over the circle map
\begin{equation}
    u_{k+1}=(u_k+\rho)\mod 1,\;u_0\in [0,1]
    \label{equ1}
\end{equation}
with the return time $T(u)$. Starting at $u=u_0$, the particle  \blue{follows $n$ iterations} of the circle map within the time
\begin{equation}
    t(u_0)=\sum_{k=0}^{n-1}T(u_k).
    \label{eq5.1}
\end{equation}
Convergence of the sum (\ref{eq5.1}) to the integral
\begin{equation}
\langle t\rangle=n\int_0^1T(u)du=n\langle T\rangle,
\label{eq5.1a}
\end{equation}
was studied in \cite{DLPS}. The sum converges, 
except for rational and a special class of irrational numbers $\rho$. 
We are interested in finding the variance of the distribution of
$t(u_0)$ around the mean value (\ref{eq5.1a}) in the case where $T(u)$ has a singularity.

As we have seen in the previous section, the inverse characteristics, 
the dependence of the variance of the number of steps on time,
is the combination of the \textit{trend}, which determines 
its large-scale asymptotic behavior, 
and the \textit{decoration} that describes fluctuations on a short time scale. 
While the decoration is governed by the continued fraction representation 
of the rotation number $\rho$, the trend is determined by the singularity 
of the return time dependence $T(u)$.

The goal of the present section is the quantitative evaluation
of the trends by means of simplified models.

\subsection{Weak power-law singularities}

First, we consider the case of a weak power-law singularity
of the return time,

\begin{equation}
T(u)=\frac{1}{|u-1/2|^a},\quad 0<a<1/2.
\label{eq5.0}
\end{equation}
In that case,
\begin{equation}
\langle T\rangle=\int_0^1T(u)du=\frac{2^a}{1-a},
\label{eq5.0a}
\end{equation}
Since we are interested in finding the trend
 of the variance growth, \blue{and assume that this trend}
 is independent of $\rho$, 
 we estimate the variance of the sum (\ref{eq5.1}) \blue{by} replacing the actual \blue{$\rho$-dependent} distribution of points with an \textit{equidistant} sequence
 \begin{equation}
u_k=u_0+\frac{k}{n},\;0\leq k\leq n-1,
\label{eq5.0b}
\end{equation}
where the initial point $u_0=c/n$, $0\leq c\leq 1$, is arbitrary.
\green{With (\ref{eq5.0}) and (\ref{eq5.0b}), the sum (\ref{eq5.1}) can be calculated in terms of the Hurwitz $\zeta$-function, $\zeta(a,c)$, (see Appendix \ref{app:hurwitz}) as}  
\begin{equation}
t(c)=\langle T\rangle n+k(c)n^a+o(n^a),\; n\gg 1,
\label{eq5.4}
\end{equation}
where
$$k(c)=\zeta(a,c)+\zeta(a,1-c).$$

Assume a uniform distribution for $c$ and calculate the
statistical properties of the time distribution. Note that
$$\delta t(c)=t-\langle t\rangle=n^a[\zeta(a,c)+\zeta(a,1-c)]$$
and
\begin{equation}
\langle\delta
t(c)\rangle_c=2n^a\int_0^{1/2}[\zeta(a,c)+\zeta(a,1-c)]dc=
2n^a\int_0^1\zeta(a,c)dc=0.
\label{eq5.5}
\end{equation}
The 
variance of $t$ for a fixed $n$ is
$$\langle(\delta
t(c))^2)\rangle_c=C(a)n^{2a},$$
where
\begin{equation}
C(a)=\langle k^2(s)\rangle=2\int_0^{1/2}[\zeta(a,c)+\zeta(a,1-c)]^2dc.
\label{eq5.6}
\end{equation}
Because $\zeta(a,c)\sim 1/c^a$ as $c\to 0$, integral (\ref{eq5.6})
converges for $a<1/2$.

Inverting (\ref{eq5.4}) at large $n$, $t$, we obtain 
$$n=\frac{t}{\langle T\rangle}-\frac{k(c)}{\langle
T\rangle}\left(\frac{t}{\langle T\rangle}\right)^a.$$
Taking into account (\ref{eq5.5}), we find that
$$\langle n\rangle_c=\frac{t}{\langle T\rangle},$$
thus the variance of the number of steps
\begin{equation}
(\Delta n(t))^2=\langle(n-\langle n\rangle_c)^2\rangle_c=K(a)t^{2a},
\label{eq5.7}
\end{equation}
where
$$K(a)=\frac{C(a)}{\langle
T\rangle^{2(a+1)}}.$$


\red{We emphasize} that \red{the} estimate
can be applied only if $(\Delta n)^2\gg 1$. We expect that formula (\ref{eq5.7}) obtained for the particular return-time function (\ref{eq5.0}), is valid for other dependences $T(u)$ with the same kind of singularity.

\red{
Expression (\ref{eq5.7}) describes correctly the anomalous asymptotic law of {variance}.} 
However, because of the equidistant
distribution assumption done above, it does not reflect
the \red{nontrivial ``decoration" of the variance law depending} on the rotation number $\rho$. 

For $a\geq 1/2$, integral (\ref{eq5.6}) diverges, therefore the estimate
(\ref{eq5.7}) is not valid. For the description of that case, another model is suggested (see \green{Appendix \ref{app:superdiffusion}}).

\subsection{\red{Asymmetric}
logarithmic singularity}

\red{As the example of} an asymmetric logarithmic 
singularity \red{of the return time, we take}
$$T(x)=1-\ln\left(\frac{1}{2}-x\right),\;x<\frac{1}{2};\;T(x)=
1-\ln 2-2\ln\left(x-\frac{1}{2}\right),\;x>\frac{1}{2},$$
\red{thus}
\begin{equation}
\langle T\rangle=\frac{5}{2}+\ln 2.
\label{eq5.8}
\end{equation}

\red{The problem can be treated in a way similar to that described in the previous subsection. Straightforward but tedious computations presented in Appendix~\ref{derivation_asym} give the following formula:}
\begin{equation}
(\Delta n)^2=\frac{1}{\langle T\rangle^2}\left[\frac{1}{12}\ln^2\left(\frac{t}{\langle 
T\rangle}\right)+0.382\ln\left(\frac{t}{\langle
T\rangle}\right)+2.36\right]+O\left(\frac{\ln^2t}{t}\right).
\label{eq5.11}
\end{equation}





\subsection{Symmetric logarithmic singularity}
In the case of a symmetric logarithmic singularity,
\begin{equation}
T(x)=1-\ln|x-1/2|,
\label{eq22a}
\end{equation}
we find
$$\langle T\rangle=2+\ln 2.$$
\red{Computations similar to those done in the previous subsections give a closed expression for the limit variance of the number of steps that}
does not grow with $t$ (see Appendix~\ref{derivation_symm}). But our approximation is based on the assumption $(\Delta n)^2\gg 1$, \red{which is not satisfied in this case}. Thus, the applied approach is not self-consistent in the case of a symmetric logarithmic singularity.

\subsection{Stronger power law singularities}
\blue{Here, we treat the case when the power in the growth law
near the singularity equals or exceeds 1/2. This situation
is modeled by the special flow with the function}
\begin{equation}
T(x)=1/|x-1/2|^a,\;1/2\leq a<1.
\label{eq5.2.1}
\end{equation}
\blue{The above approach is not applicable here due to absence of
convergence in the integral} \green{(\ref{eq5.6}).}

\subsubsection{CTRW model}
In this situation, we apply another simplification of the original problem.
Assume that a particle stays at the point $x$, $0<x<1$, during the time
interval (\ref{eq5.2.1}),
and then jumps instantly to another point of the same interval, chosen
randomly with the same probability for all destination points. 
\blue{In this way, we replace the circle map by the random map of the
unit interval.}
\red{Hence, we ignore the non-random details of the walk (\ref{equ1}) determined by the rotation number $\rho$.}
The process is a unidirectional CTRW \cite{KlSo}.

\red{
Assume that the initial position of the particle, $x_0$, is uniformly distributed in the interval $[0,1]$; then it remains uniformly distributed after any number of steps. According to the law (\ref{eq5.2.1}), the particle cannot jump at $t<2^a$, and it \green{jumps} after the time interval $t>2^a$, if its coordinate is $x_+(t)=1/2+1/t^a$ or $x_-(t)=1/2-1/t^a$. 
Therefore, the probability that the particle \green{jumps} during the time interval $[t,t+dt]$ is $(|dx_+/dt|+|dx_-/dt|)dt$.} 
Thus, for the waiting time, we obtain: 
\begin{equation}
w(t)=\left\{\begin{array}{ll}
0,& t<2^a\\[1ex]
\displaystyle\frac{2}{a}\,t^{-1-1/a},& t>2^a
\end{array}.\right.
\label{eq5.2.3}
\end{equation}
Because the second moment $$\int_0^{\infty}t^2w(t)dt$$
diverges for $1/2\leq a<1$, we expect that the dispersion of
particles is anomalous in that case.   
 
To proceed, \red{let us} calculate the Laplace transform of the waiting time pdf (\ref{eq5.2.3}). The Laplace transform of (\ref{eq5.2.3}),
\begin{equation}
\tilde{w}(s)=\int_{2^a}^{\infty}e^{-st}\frac{2}{a}t^{-1/a-1}dt
\label{eq5.2.4}
\end{equation}
is reduced to the incomplete $\Gamma$ function by transformation
$u=st$:
$$\tilde{w}(s)=\frac{2}{a}s^{1/a}\Gamma\left(-\frac{1}{a},2^as\right).$$
In order to analyze the transport at large $t$, we need the
asymptotics of $\tilde{w}(s)$ at small $s$.

For $1<1/a<2$, we apply the integration by parts \green{twice} to \green{the} integral
(\ref{eq5.2.4}) until we get the
incomplete $\Gamma$ function with the
first argument equal to $2-1/a$, which is positive, and use the
standard expansion
\begin{equation}
\Gamma(q,z)=  
\Gamma(q)-e^{-z}z^p\sum_{k=0}^{\infty}\frac{z^k}{q(q+1)\ldots(q+k)}
\label{eq5.2.5}
\end{equation}
valid for $q>0$. We
obtain the following expression for $\tilde{w}(s)$:
$$\tilde{w}(s)=
\frac{2}{a}s^{1/a}\exp(-2^as)
\left[\Gamma(-1/a)-\sum_{k=0}^{\infty}\frac{(2^as)^{-1/a+m}}
{(-1/a)(-1/a+1)\ldots(-1/a+m)}\right]\sim$$
\begin{equation}
1-As+Bs^{1/a}-Cs^2+\ldots;
\label{eq5.2.7}
\end{equation}
\begin{equation}
A=\frac{2^a}{1-a},\;B=-2\Gamma(1-1/a),\;C=\frac{2^{2a-1}}{2a-1}.
\label{eq5.2.8}
\end{equation} 
In the \green{important} limit $a\to 1/2$, one obtains
\begin{equation}
\tilde{w}(s)=1-\bar{A}s-Ds^2\ln(s)+Es^2+\ldots;
\label{eq5.2.8a}
\end{equation}
\begin{equation}
\bar{A}=2\sqrt{2},\;D=2,\;E=3-2\gamma-\ln(2),
\label{eq5.2.9}
\end{equation}
where $\gamma$ is the Euler constant.

\subsubsection{Dispersion of particles}

To analyze the dispersion of particles, we find the
probability distribution $\chi_n(t)$ of
the random variable $n(t)$, the number of steps $n$ 
performed during the time $t$, and calculate the momenta
$$I_m(t)=\sum_{n=1}^{\infty}n^m\chi_n(t).$$
According to the general theory of the CTRW (see, e.g., \cite{KlSo}),
the Laplace transform of function \red{$\chi_n(t)$} is given by the formula
\begin{equation}
\tilde{\chi}_n(s)=\tilde{w}^n(s)\frac{1-
\tilde{w}(s)}{s}.
\label{eq5.2.10}
\end{equation}
The Laplace transform of the $m$-th moment is
\begin{equation}
\tilde{I}_m(s)=\sum_{n=1}^{\infty}n^m\tilde{\chi}_n(s).
\label{eq5.2.11}
\end{equation}

\red{Calculating the sums and taking the inverse Laplace transform, we obtain the following expressions (see Appendix \ref{app:superdiffusion}).

For $1/2<a<1$,}
\begin{equation}
r_2=\langle(n-\langle
n\rangle)^2\rangle=\red{\langle n^2\rangle-(\langle n\rangle)^2}=\frac{a^2(1-a)^3}{2^{3a-2}(2a-1)(3a-1)}t^{3-1/a}-\frac{a^2(1-a)}{2^a(2a-1)}t+\ldots,
\label{eq5.2.12}
\end{equation}
\begin{equation}
r_4^2\equiv\langle(n-\langle
n\rangle)^4\rangle=\red{2(\langle n^4\rangle)^2-8\langle n\rangle\langle n^3\rangle+6(\langle n^2\rangle)^2}
\sim t^{5-1/a}2^{4-5a}\frac{a^2(1-a)^5}{(4a-1)(5a-1)},
\label{eq5.2.18}
\end{equation}   
\begin{equation}
\langle(n-\langle n\rangle)^m\rangle\sim K_mt^{m+1-1/a},\;m>4,
\label{eq5.2.22}
\end{equation}
where $K_m$ are constant \blue{factors}.

In the limit $a\to 1/2$, we obtain \green{the following}:
\begin{equation}
r_2=\frac{\sqrt{2}}{8}t\ln(t)-\frac{5+\ln(2)}{8\sqrt{2}}t+\ldots,
\label{eq5.2.14a}
\end{equation}
\begin{equation}
r_4^2\sim t^{3}\frac{\sqrt{2}}{96}.
\label{eq5.2.21}
\end{equation}
Note that the latter formula \red{can be} obtained from (\ref{eq5.2.18}) in the
limit $a\to 1/2$.

Formulas (\ref{eq5.2.14a}), (\ref{eq5.2.18}) and (\ref{eq5.2.22})
\blue{unequivocally demonstrate that the growth law of the fourth central moment 
of the distribution of particles is different from the growth law of the squared 
second central moment; this fact allows us}
to conclude that dispersion of particles is a multifractal process.


\section{Conclusions and Discussion}
\label{sect:conclusions}

\red{We have considered the deterministic transport of tracers in stationary, two-dimensional, spatially periodic flows with different kinds of stagnation points. The relation between transport anomalies (subdiffusion and superdiffusion) and the types of stagnation points has been studied by direct numerical simulations of the tracer dynamics. It has been found that, while the details of the dependence of the particle distribution variance are strongly influenced by the rotation number of the flow, the large-time asymptotic trends of that dependence are determined solely by the passage-time singularity near the stagnation point. The transport characteristics are reproduced by the simulation of appropriate special flows. The large-time trends of particle transport are described in the framework of simplified models of special flows, namely the equidistant point model in the case of subdiffusion and the CTRW model in the case of superdiffusion.} 
 \appendix
\section{Best approximations to the irrational rotation numbers employed in Fig.~\ref{fig:decorations}}
\label {app:approx}
Given an irrational $\rho$, the rational number $m/n$ which minimizes the distance 
$|\rho-m/n|$ among all rationals with denominators that do not exceed $n$
is traditionally called the best approximation to $\rho$. A sequence of best approximations is obtained 
by truncating the expansion of $\rho$ into the infinite continued fraction.

\begin{itemize}
\item[(a)] For the golden mean $(\sqrt{5}-1)/2$, the best approximations 
are $F_n/F_{n+1}$ where $F_n$ is the $n$-th Fibonacci number: \ 
$\displaystyle\frac{1}{2},\frac{2}{3},\frac{3}{5},\frac{5}{8},\frac{8}{13},\frac{13}{21},\frac{21}{34},\frac{34}{55},\frac{55}{89},\frac{89}{144},\frac{144}{233},
\frac{233}{377},\frac{377}{610},\frac{610}{987},\frac{987}{1597},\frac{1597}{2584},\frac{2584}{4181},
\ldots$
\item[(b)]The best rational approximations to $(\sqrt{3}-1)/2$ are:\\
$\displaystyle\frac{1}{3},\frac{3}{8},\frac{4}{11},\frac{11}{30},\frac{15}{41},\frac{41}{112},
\frac{56}{153},\frac{153}{418},\frac{209}{571},\frac{571}{1560},\frac{780}{2131},
\frac{2131}{5822},\frac{2911}{7953},\frac{7953}{21728},\frac{10864}{29681},\ldots$
\item[(c)]The best rational approximations to $1/\mbox{e}$ are:\\
$\displaystyle\frac{3}{8},\frac{4}{11},\frac{4}{11},\frac{7}{19},\frac{32}{87},\frac{39}{106},
\frac{71}{193},\frac{465}{1264},\frac{536}{1457},\frac{1001}{2721},\frac{8544}{23325},
\frac{9545}{25946},\ldots$
\item[(d)] The best rational approximations to $\sqrt{2}-1$ are:\\
$\displaystyle\frac{1}{2},\frac{2}{5},\frac{5}{12},\frac{12}{29},\frac{29}{70},\frac{70}{169},
\frac{169}{408},\frac{408}{985},\frac{985}{2378},\frac{2378}{5741},\frac{5741}{13860},
\frac{13860}{33461},\ldots$
\end{itemize}
\green{\section{Hurwitz $\zeta$-function and derivation of formula (\ref{eq5.4})}
\label{app:hurwitz}
The Hurwitz $\zeta$-function, $\zeta(a,c)$, is defined at real $a>1$ as 
\begin{equation}
  \zeta(a,c)=\sum_{k=0}^{\infty}\frac{1}{(k+c)^a}.
  \label{A1}
  \end{equation}
  Then the finite sum
  \begin{equation}
      \sum_{k=0}^{n-1}\frac{1}{(k+c)^a}=\zeta(a,c)-
  \sum_{k=n}^{\infty}\frac{1}{(k+c)^a}.
  \label{A2}
  \end{equation}
  The asymptotics of the second term in the right-hand side of (\ref{A2}) can be found by the replacement of the sum with the integral,
  $$\sum_{k=n}^{\infty}\frac{1}{(k+c)^a}\sim \int_{n}^{\infty}\frac{dk}{(k+c)^a}=\frac{n^{1-a}}{a-1}.$$
  Thus,
  \begin{equation}
      \sum_{k=0}^{n-1}\frac{1}{(k+c)^a}\sim\frac{n^{1-a}}{a-1}+\zeta(a,c).
      \label{A3}
  \end{equation} 

  The Hurwitz $\zeta$-function is known to be a meromorphic function of $a$ with the only singularity at $a=1$ \cite{Hurwitz}. Therefore, the relation (\ref{A3}) can be extended to the region $a<1$ and used to evaluate expression (\ref{eq5.1}). Let us present that expression with (\ref{eq5.0}) and (\ref{eq5.0b}) as a sum of two expressions corresponding to the points to the left and to the right of the singularity point $x=1/2$:
  $$t(c)=\sum_{k=0}^{n/2-1}\frac{1}{\left(\frac{1}{2}-\frac{k+c}{n}\right)^a}+\sum_{k=n/2}^{n-1} \frac{1}{\left(\frac{k+c}{n}-\frac{1}{2}\right)^a}.$$
  Denoting $l=n/2-k-1$ for the first sum and $l=k-n/2$ for the second sum, we find
  $$t(c)=\sum_{l=0}^{n/2-1}\frac{n^a}{(l+1-c)^a}+\sum_{l=0}^{n/2-1}\frac{n^a}{(l+c)^a}.$$
  Using the asymptotic formula (\ref{A3}), we find:
  $$t(c)\sim n\frac{2^a}{1-a}+\zeta(a,c)+\zeta(a,1-c).$$
  Taking into account expression (\ref{eq5.0a}, we obtain (\ref{eq5.4}).
}

\red{\section{Asymmetric logarithmic singularity: Derivation of formula (\ref{eq5.11})}
\label{derivation_asym}       
In the case of an asymmetric logarithmic singularity and equidistant point distribution, 
$$x_k=(c+k)/n,\;0\leq k\leq n-1,$$
formula (\ref{eq5.1}) can be written as}
$$t(c)=-\sum_{l=0}^{n/2-1}\ln(l+1-c)-2\sum_{l=0}^{n/2-1}\ln(l+c)+\frac{3}{2}n\ln(n)+n\left(1-\frac{\ln 2}{2}\right).$$
Taking into account that
\begin{equation}
\sum_{l=0}^{n/2-1}\ln(l+c)=\ln\prod_{l=0}^{n/2-1}(l+c)=\ln\frac{\Gamma(n/2+c)}{\Gamma(c)},
\label{eq5.8a}
\end{equation}
\begin{equation}
\sum_{l=0}^{n/2-1}\ln(l+1-c)=\ln\frac{\Gamma(n/2+1-c)}{\Gamma(1-c)}
\label{eq5.8b}
\end{equation}
and using the asymptotic formula
\begin{equation}
\ln\Gamma(z)=z\ln z-z-\frac{1}{2}\ln z+\frac{1}{2}\ln(2\pi))+o(1),
\label{eq5.8c}
\end{equation}
we obtain the following asymptotic formula for $t(c)$:
\begin{equation}
t(c)=\langle T\rangle n+\left(\frac{1}{2}-c\right)\ln n+k(c),
\label{eq5.9}
\end{equation}
where
$$k(c)=(c-2)\ln 2-\frac{3}{2}\ln\pi+\ln\Gamma(1-c)+2\ln\Gamma(c).$$
Note that $\langle(1/2-c)\rangle_c=0$, $\langle k(c)\rangle_c=0$, hence $\langle t(c)\rangle_c=\langle T\rangle n$.

Inverting (\ref{eq5.9}) at large $n$, $t$, we obtain:
\begin{equation}
n(c)=\left[t+\left(c-\frac{1}{2}\right)\ln\left(\frac{t}{\langle T\rangle}\right)-k(c)\right]/\langle T\rangle+O\left(\frac{\ln(t)}{t}\right),
\label{eq5.10}
\end{equation}
where $\langle T\rangle$ is determined by relation (\ref{eq5.8}). Thus,
$$(\Delta n(c))^2=\frac{1}{\langle T\rangle^2}\left[\left(c-\frac{1}{2}\right)\ln\left(\frac{t}{\langle 
T\rangle}\right)-k(c)\right]^2+O\left(\frac{\ln^2t}{t}\right).$$
Averaging over $c$, we find
\begin{equation}
(\Delta n)^2=\frac{1}{\langle T\rangle^2}\left[\frac{1}{12}\ln^2\left(\frac{t}{\langle 
T\rangle}\right)+0.382\ln\left(\frac{t}{\langle
T\rangle}\right)+2.36\right]+O\left(\frac{\ln^2t}{t}\right).
\label{eq5.10a}
\end{equation}

\section{Variance in the case of a symmetric logarithmic singularity}
\label{derivation_symm}
In the case 
$$T(x_k)=1-\ln|x_k-1/2|,\;x_k=(c+k)/n$$
we get
$$t(c)=n+n\ln n-\sum_{l=0}^{n/2-1}\ln(l+1-c)-\sum_{l=0}^{n/2-1}\ln(l+c).$$
Using relations (\ref{eq5.8a})-(\ref{eq5.8c}), we obtain
$$t(c)=n\langle T\rangle+\tilde{k}(c)+o(1),$$
where
$$\tilde{k}(c)=1-\ln(2\pi)+\ln\Gamma(1-c)+\ln\Gamma(c).$$
Thus,
$$n(c)=\frac{t-\tilde{k}(c)}{\langle T\rangle},\;
(\Delta n(c))^2=\frac{\tilde{k}(c)^2}{\langle T\rangle^2}+o(1),$$
and the variance
$$(\Delta n)^2=\langle\Delta n(c)^2\rangle_c=\frac{
\langle\tilde{k}(c)^2\rangle_c}{\langle T\rangle^2}+o(1)$$
does not grow with $t$.

\section{Derivation of formulas describing dispersion of particles  
in the case of superdiffusion}
\label{app:superdiffusion}
\red{Let $1/2<a<1$. Using expressions (\ref{eq5.2.7}), (\ref{eq5.2.10}) and} formulas
$$\sum_{n=1}^{\infty}nu^n=\frac{u}{(1-u)^2},\qquad
\sum_{n=1}^{\infty}n^2u^n=\frac{u(1+u)}{(1-u)^3},$$ 
$$\sum_{n=1}^{\infty}n^3u_n=\frac{u(1+4u+u^2)}{(1-u)^4},\qquad
\sum_{n=1}^{\infty}n^4u_n=\frac{u(1+11u+11u^2+u^3)}{(1-u)^5},$$
\red{we calculate the sums in (\ref{eq5.2.11}):
$$\tilde{I}_1(s)=\frac{1}{As^2}+\frac{B}{A^2}s^{1/a-3}-\left(1+\frac{C}{A^2}\right)\frac{1}{s}+\ldots,$$}
$$\tilde{I}_2(s)=
\frac{2}{A^2s^3}+\frac{4B}{A^3}s^{1/a-4}-\frac{2}{A^2}\left(\frac{3A}{2}+\frac{2C}{A}\right)
\frac{1}{s^2}+\ldots,$$
$$\tilde{I}_3(s)=\frac{6}{A^3s^4}+\frac{18B}{A^4}s^{1/a-5}+\ldots,\;
\tilde{I}_4(s)=\frac{24}{A^4s^5}+\frac{96B}{A^5}s^{1/a-6}+\ldots.$$
The inverse Laplace transform gives
\begin{equation}
I_1(t)=\langle
n(t)\rangle=\frac{t}{A}+\frac{B}{A^2}\frac{t^{2-1/a}}{\Gamma(3-1/a)}-\left(1+\frac{C}{A^2}\right)+
\ldots,
\label{eq5.2.11a}
\end{equation}
\begin{equation}
I_2(t)=\langle
n^2(t)\rangle=\frac{t^2}{A^2}+\frac{4B}{A^3}\frac{t^{3-1/a}}{\Gamma(4-1/a)}
-\frac{3A^2+4C}{A^3}t+\ldots,
\label{eq5.2.11b} 
\end{equation}  
\begin{equation}
I_3(t)=
\frac{t^3}{A^3}+\frac{18B}{A^4}\frac{t^{4-1/a}}{\Gamma(5-1/a)}+\ldots,\;
I_4(t)=\frac{t^4}{A^4}+\frac{96B}{A^5}\frac{t^{5-1/a}}
{\Gamma(6-1/a)}+\ldots
\label{eq5.2.17}
\end{equation}

Hence, the variance
$$r_2=\langle(n-\langle
n\rangle)^2\rangle=I_2(t)-I_1^2(t)=\frac{B}{A^3}
\left(\frac{4}{\Gamma(4-1/a)}-\frac{2}{\Gamma(3-1/a)}\right)t^{3-1/a}-\frac{A^2+2C}{A^3}t.$$
Substituting (\ref{eq5.2.8}), we find
\begin{equation}
\frac{a^2(1-a)^3}{2^{3a-2}(2a-1)(3a-1)}t^{3-1/a}-\frac{a^2(1-a)}{2^a(2a-1)}t+\ldots.
\label{eq5.2.12a}
\end{equation}
\red{For} the quantity
\begin{equation}
r_4^2\equiv\langle(x-\langle
x\rangle)^4\rangle=2I_4^2-8I_1I_3+6I_2^2,
\label{eq5.2.15}
\end{equation}
\red{using} (\ref{eq5.2.11a}), (\ref{eq5.2.11b})
and
(\ref{eq5.2.17}), we
find
\begin{equation}
r_4^2\sim t^{5-1/a}2^{4-5a}\frac{a^2(1-a)^5}{(4a-1)(5a-1)}.
\end{equation}   
\red{Similarly, we find that for} $m>4$,
$$
I_m(t)=\left(\frac{t}{2\sqrt{2}}\right)^m+C_mt^{m+1-1/a}+\ldots,$$
which allows to conclude that
\begin{equation}
\langle(n-\langle n\rangle)^m\rangle\sim K_mt^{m+1-1/a},
\label{eq5.2.22a}
\end{equation}
where $C_m$ and $K_m$ are constant.

In the limit $a\to 1/2$, we obtain:
$$\tilde{I}_1(s)=\frac{1}{2\sqrt{2}}\cdot\frac{1}{s^2}-\frac{5+2\gamma+
\ln(2)}{8}\cdot\frac{1}{s}-\frac{1}{4}\frac{\ln(s)}{s}+\ldots,$$
$$\tilde{I}_2(s)=\frac{1}{4s^3}-\frac{12+8\gamma+4\sqrt{2}}{16\sqrt{2}}
\frac{1}{s^2}-\frac{\ln(s)}{s^2}+\ldots,$$
$$\tilde{I}_3(s)=
\frac{6}{A^3s^4}-\frac{18D}{A^4}\frac{\ln(s)}{s^3}+
\frac{6(-2A^2+3E)}{A^4s^3}+\ldots,
$$
$$\tilde{I}_4(s)=\frac{24}{A^4s^5}-\frac{96D}{A^5}\frac{\ln(s)}{s^4}+\frac{12(-5A^2+8E)}{A^5s^4}+
\ldots, $$
hence
\begin{equation}
I_1(t)=\frac{1}{2\sqrt{2}}t+\frac{1}{4}\ln(t)-\frac{5+\ln(2)}{8}+\ldots,
\label{eq5.2.13}
\end{equation}
\begin{equation}
I_2(t)=\frac{t^2}{8}+\frac{\sqrt{2}}{4}t\ln(t)
-\frac{5+\ln(2)}{4\sqrt{2}}t+\ldots,
\label{eq5.2.14}
\end{equation}  
\red{and} the variance
\begin{equation}
r_2=\frac{\sqrt{2}}{8}t\ln(t)-\frac{5+\ln(2)}{8\sqrt{2}}t+\ldots,
\label{eq5.2.14b}
\end{equation}
\begin{equation}
I_3(t)=\frac{\sqrt{2}}{32}t^3+\frac{9}{32}t^2\ln(t)-\frac{48+9\ln(2)}{64}+\ldots,
\label{eq5.2.19}
\end{equation}  
\begin{equation}
I_4(t)=\frac{t^4}{64}+\frac{\sqrt{2}}{8}t^3\ln(t)-\frac{17+3\ln(2)}{24\sqrt{2}}t^3+\ldots,
\label{eq5.2.20}
\end{equation}
\begin{equation}
r_4^2\sim t^{3}\frac{\sqrt{2}}{96}.
\label{eq5.2.21a}
\end{equation}

ACKNOWLEDGMENT: research of MZ was carried out with the support of the DFG (project ZA 658/3-1).

\end{document}